\documentclass[onecolumn]{aastex631}
\usepackage{graphicx}

\usepackage{float}
\usepackage{amsmath}
\usepackage{amssymb}
\usepackage{mathtools}
\usepackage{mathrsfs}
\usepackage{bm}
\usepackage{enumerate}
\usepackage{booktabs}

\newcommand{\MultiNest}{\textsc{MultiNest}}
\newcommand{\Msun}{$M_\odot$}
\newcommand{\ntwolo}{N$^2$LO}
\newcommand{\nthreelo}{N$^3$LO}

\usepackage{subfiles} 

\begin{document}

\title{Constraining the dense matter equation of state with new NICER mass-radius measurements and new chiral effective field theory inputs}

\author[0000-0002-9626-7257]{Nathan Rutherford}
\email{nathan.rutherford@unh.edu; melissa.mendes@physik.tu-darmstadt.de; isak.svensson@physik.tu-darmstadt.de}
\affiliation{Department of Physics and Astronomy, University of New Hampshire, Durham, New Hampshire 03824, USA}

\author[0000-0002-5250-0723]{Melissa Mendes}
\affil{Technische Universit\"at Darmstadt, Department of Physics, 64289 Darmstadt, Germany}
\affil{ExtreMe Matter Institute EMMI, GSI Helmholtzzentrum f\"ur Schwerionenforschung GmbH, 64291 Darmstadt, Germany}
\affil{Max-Planck-Institut f\"ur Kernphysik, Saupfercheckweg 1, 69117 Heidelberg, Germany}

\author[0000-0002-9211-5555]{Isak Svensson}
\affil{Technische Universit\"at Darmstadt, Department of Physics, 64289 Darmstadt, Germany}
\affil{ExtreMe Matter Institute EMMI, GSI Helmholtzzentrum f\"ur Schwerionenforschung GmbH, 64291 Darmstadt, Germany}
\affil{Max-Planck-Institut f\"ur Kernphysik, Saupfercheckweg 1, 69117 Heidelberg, Germany}

\author[0000-0001-8027-4076]{Achim~Schwenk}
\affil{Technische Universit\"at Darmstadt, Department of Physics, 64289 Darmstadt, Germany}
\affil{ExtreMe Matter Institute EMMI, GSI Helmholtzzentrum f\"ur Schwerionenforschung GmbH, 64291 Darmstadt, Germany}
\affil{Max-Planck-Institut f\"ur Kernphysik, Saupfercheckweg 1, 69117 Heidelberg, Germany}

\author[0000-0002-1009-2354]{Anna~L.~Watts}
\affiliation{Anton Pannekoek Institute for Astronomy, University of Amsterdam, Science Park 904, 1098XH Amsterdam, the Netherlands}

\author[0000-0003-0640-1801]{Kai Hebeler}
\affil{Technische Universit\"at Darmstadt, Department of Physics, 64289 Darmstadt, Germany}
\affil{ExtreMe Matter Institute EMMI, GSI Helmholtzzentrum f\"ur Schwerionenforschung GmbH, 64291 Darmstadt, Germany}
\affil{Max-Planck-Institut f\"ur Kernphysik, Saupfercheckweg 1, 69117 Heidelberg, Germany}

\author [0000-0003-2017-4158]{Jonas Keller}
\affil{Technische Universit\"at Darmstadt, Department of Physics, 64289 Darmstadt, Germany}
\affil{ExtreMe Matter Institute EMMI, GSI Helmholtzzentrum f\"ur Schwerionenforschung GmbH, 64291 Darmstadt, Germany}

\author[0000-0002-6742-4532]{Chanda Prescod-Weinstein}
\affiliation{Department of Physics and Astronomy, University of New Hampshire, Durham, New Hampshire 03824, USA}

\author[0000-0002-2651-5286]{Devarshi~Choudhury}
\affil{Anton Pannekoek Institute for Astronomy, University of Amsterdam, Science Park 904, 1098XH Amsterdam, the Netherlands}

\author[0000-0002-9397-786X]{Geert~Raaijmakers}
\affil{GRAPPA, Anton Pannekoek Institute for Astronomy and Institute of High-Energy Physics, University of Amsterdam, Science Park 904, 1098 XH Amsterdam, the Netherlands}

\author[0000-0001-6356-125X ]{Tuomo~Salmi}
\affil{Anton Pannekoek Institute for Astronomy, University of Amsterdam, Science Park 904, 1098XH Amsterdam, the Netherlands}

\author[0009-0003-2793-1569]{Patrick Timmerman}
\affiliation{Anton Pannekoek Institute for Astronomy, University of Amsterdam, Science Park 904, 1098XH Amsterdam, the Netherlands}

\author[0000-0003-3068-6974]{Serena~Vinciguerra}
\affil{Anton Pannekoek Institute for Astronomy, University of Amsterdam, Science Park 904, 1098XH Amsterdam, the Netherlands}

\author[0000-0002-6449-106X]{Sebastien~Guillot}
\affil{IRAP, CNRS, 9 avenue du Colonel Roche, BP 44346, F-31028 Toulouse Cedex 4, France}
\affil{Universit\'{e} de Toulouse, CNES, UPS-OMP, F-31028 Toulouse, France.}

\author[0000-0002-5907-4552]{James M.~Lattimer}
\affil{Department of Physics and Astronomy, Stony Brook University, Stony Brook, NY 11794-3800, USA}

\begin{abstract}
Pulse profile modeling of X-ray data from the Neutron Star Interior Composition Explorer is now enabling precision inference of neutron star mass and radius. Combined with nuclear physics constraints from chiral effective field theory ($\chi$EFT), and masses and tidal deformabilities inferred from gravitational wave detections of binary neutron star mergers, this has led to a steady improvement in our understanding of the dense matter equation of state (EOS). Here, we consider the impact of several new results: the radius measurement for the 1.42\,\Msun\ pulsar PSR~J0437$-$4715 presented by \citet{Choudhury24}, updates to the masses and radii of PSR J0740$+$6620 and PSR J0030$+$0451, and new $\chi$EFT results for neutron star matter up to 1.5 times nuclear saturation density. Using two different high-density EOS extensions---a piecewise-polytropic (PP) model and a model based on the speed of sound in a neutron star (CS)---we find the radius of a 1.4\,\Msun\ (2.0\,\Msun) neutron star to be constrained to the 95\% credible ranges $12.28^{+0.50}_{-0.76}\,$km ($12.33^{+0.70}_{-1.34}\,$km) for the PP model and $12.01^{+0.56}_{-0.75}\,$km ($11.55^{+0.94}_{-1.09}\,$km) for the CS model. The maximum neutron star mass is predicted to be $2.15^{+0.14}_{-0.16}\,$\Msun\ and $2.08^{+0.28}_{-0.16}\,$\Msun\ for the PP and CS models, respectively. We explore the sensitivity of our results to different orders and different densities up to which $\chi$EFT is used, and show how the astrophysical observations provide constraints for the pressure at intermediate densities. Moreover, we investigate the difference $R_{2.0} - R_{1.4}$ of the radius of 2\,\Msun\ and 1.4\,\Msun\ neutron stars within our EOS inference.
\end{abstract}


\section{Introduction}
\label{sec:intro}

The increasingly precise measurement of neutron star properties such as mass, radius, and tidal deformability, enabled by new observational facilities and techniques, informs our understanding of the equation of state (EOS) of supranuclear density matter. Radio timing measurements of high pulsar masses \citep[][]{Antoniadis13,Arzoumanian18,Cromartie20,Fonseca21,Shamohammadi23} and gravitational wave (GW) measurements of tidal deformability from neutron star binary mergers \citep[][]{gw170817,gw190425} have now been supplemented by measurements of neutron star mass and radius for X-ray pulsars using data from the Neutron Star Interior Composition Explorer \citep[NICER;][]{Gendreau16}. These astrophysical measurements have been used in various analyses, often in combination with constraints from nuclear theory and laboratory experiments, to place limits on the properties of neutron-rich matter, possible quark or hyperon phases in neutron star cores, and the presence of dark matter in and around neutron stars \citep[see, e.g.,][]{Miller21,Raaijmakers21,Legred21,Biswas22,Huth22,Miao_2022,Giangrandi2023,Rutherford23,SunX23,Takatsy23,Annala23,Pang24,Koehn24,Kurkela24,Shakeri2024}. 

Pulse profile modeling (PPM), the relativistic ray-tracing-based inference technique used to derive masses and radii from NICER data, is applied to X-ray-bright rotation-powered millisecond pulsars (MSPs). Full details of the PPM process can be found in \citet{Bogdanov19b,Bogdanov21}. So far, mass-radius inferences have been published for the MSPs PSR J0030$+$0451 \citep[hereafter, J0030;][]{Miller19,Riley19} and PSR J0740$+$6620 \citep[hereafter, J0740;][]{Miller21,Riley21}. J0030 is an isolated pulsar for which there is no independent constraint on the mass; J0740 is in a binary and the mass ($2.08 \pm 0.07\,$\Msun) is well constrained by radio pulsar timing \citep{Fonseca21}. The inferred radii for these two sources have uncertainties at the $\pm 10$ \% level [68\% credible interval (CI)]. Follow-on studies have looked more closely at specific aspects of the analysis, such as the treatment of background \citep{Salmi22}, the atmospheric model \citep{Salmi23}, and simulation and sampler resolution settings \citep{Vinciguerra23}. 

NICER data have now enabled inference of the mass and radius for PSR J0437$-$4715 \citep[hereafter, J0437;][]{Choudhury24}, the closest and brightest MSP. This is a challenging source to model: the presence of a bright active galactic nucleus in the field of view requires the spacecraft to observe off-axis, and despite this there is still a substantial background contribution from this source. However, J0437 is also a binary MSP with a well-constrained mass from radio pulsar timing of  $M = 1.418 \pm 0.044\,$\Msun\ \citep{Reardon24}. The tightly constrained radio-timing-derived mass (and distance and inclination) is used as a prior for the PPM analysis, and this has enabled radius constraints at the $\pm 7$\% level (68\% CI). Precise radius information for typical $1.4\,$\Msun\ neutron stars plays an important role for constraining the dense matter EOS, because this correlates well with the pressure of neutron-rich matter around twice saturation density \citep[see, e.g.,][]{Lattimer2001,Lattimer2013,Drischler2021,Lim2024} and thus provides key constraints at intermediate densities.

There are also new results for J0030 and J0740. \citet{Vinciguerra24} carried out a reanalysis of the J0030 data set from \citet{Riley19}, using an upgraded PPM pipeline and instrument response model, and incorporating background constraints. This source now appears to be more complex than first thought, with different modes (corresponding to different hot spot geometries)\footnote{The hot spots, which give rise to the pulsation as the star rotates and the thermal emission from the magnetic poles of the star, are thought to arise due to the heat generated from magnetospheric return currents \citep[see, e.g.,][]{Ruderman1975,Arzoumanian18,Harding2001,Salmi20}.} that have different inferred masses and radii. Meanwhile a larger data set for J0740 has enabled more robust constraints on the mass and radius for that source \citep{Dittmann24,Salmi24}. This is thus an opportune moment to update our EOS analyses.

In parallel to these astrophysical advances, there have been great developments on the EOS around nuclear densities based on chiral effective field theory ($\chi$EFT) interactions \citep[see, e.g.,][]{Epelbaum2009,Machleidt2011,Hammer2013,Hebeler2021}. Combined with powerful many-body methods, $\chi$EFT interactions have enabled calculations of neutron matter up to around nuclear saturation density ($n_0 = 0.16\,$fm$^{-3}$) that provide important constraints for the EOS of the outer core of neutron stars \citep[see, e.g.,][]{Hebeler2013,Lynn2019,Drischler2021ARNPS,Huth2021}. In our previous multimessenger analyses \citep{Raaijmakers19,Raaijmakers20,Raaijmakers21}, we have used the $\chi$EFT constraints from \citet{Hebeler2013,Tews2013,Lynn2016,Drischler2019} to explore EOS inference from the NICER results derived using the X-ray Pulse Simulation and Inference \citep[X-PSI;][]{Riley23} PPM pipeline \citep{Riley19,Riley21}\footnote{The results of \citet{Miller19,Miller21} are derived using an independent PPM pipeline.} in combination with GW-derived tidal deformabilities.

The EOS inference requires prior assumptions over all densities. To this end, we have used two different high-density EOS extensions---a piecewise-polytropic (PP) model \citep{Hebeler2013} and a model based on the speed of sound in a neutron star (CS) \citep{Greif19}---to cover the full EOS space beyond a fiducial density of $1.1 n_0$, up to which the $\chi$EFT calculations were trusted. Recently, new $\chi$EFT calculations from \citet{Keller2023} of neutron star matter in beta equilibrium and up to $1.5 n_0$ have been presented. In this work, we explore new prior EOS ensembles based on these new $\chi$EFT calculations at different chiral orders---next-to-next-to-leading order (N$^2$LO) and next-to-next-to-next-to-leading order (N$^3$LO)---as well as different transition densities ($1.1 n_0$ and $1.5 n_0$) to the PP and CS models.

This paper is organized as follows. In Sec.~\ref{sec:methods}, we introduce our Bayesian inference framework for providing constraints on the dense matter EOS and the properties of neutron stars. Sections~\ref{sec:newceft} and~\ref{sec:newpriors} discuss the new $\chi$EFT calculation and its implementation in our framework, as well as the new prior distributions. The astrophysical constraints are summarized in Sec.~\ref{sec:astro}. In addition to a ``Baseline'' scenario consisting of the GW observations from GW170817 and GW190425 \citep{gw170817,gw190425} and the previously explored NICER sources J0740 and J0030 \citep{Salmi22,Vinciguerra24}, we investigate a ``New'' scenario with the new J0437 and J0740 NICER results \citep{Choudhury24,Salmi24} and the revised analysis including background constraints for J0030 from \citet{Vinciguerra24}. Other scenarios are explored in the Appendix. In Sec.~\ref{sec:results}, we first study the changes due to the new priors for the ``Baseline'' scenario and then turn to the impact of the new observations on the dense matter EOS and the properties of neutron stars. Finally, we discuss implications and conclude in Sec.~\ref{sec:discuss}.

\section{Methodology}
\label{sec:methods}

We begin by discussing the \citet{Raaijmakers21} Bayesian inference framework, the new \ntwolo\ and \nthreelo\ $\chi$EFT calculations from \citet{Keller2023} and their implementations into this framework, the resulting mass-radius and pressure-energy density prior distributions, and the usage of the available astrophysical constraints.

\subsection{Bayesian inference framework}

In this work, we follow the analysis framework used in \citet{Raaijmakers21}, which builds on the work of \citet{Greif19,Raaijmakers19,Raaijmakers20}. Here, we briefly summarize the method and outline any modifications. We use the open-source EOS inference code NEoST\footnote{\url{https://github.com/xpsi-group/neost}} (\texttt{v0.10}) \citep{Raaijmakers24}, which implements this framework\footnote{NEoST in prerelease form was also used in \citet{Greif19,Raaijmakers19,Raaijmakers20,Raaijmakers21,Rutherford23}.}. A full reproduction package, including the posterior samples and scripts to generate the plots in this
Letter, are available in a Zenodo repository at \citet{plotdata}.  

For the high-density extension of the EOS  we consider two different parameterizations:  i) a PP model with three segments between varying transition densities \citep[][]{Hebeler2013}; and ii) a CS model first introduced in \citet{Greif19}. Below a transition density (which for our main results we take to be $1.5 n_0$, but also explore the past choice of $1.1 n_0$), these parameterizations are matched to a single polytropic fit to the EOS range calculated from $\chi$EFT interactions. The latter are discussed in more detail in Sec.~\ref{sec:newceft}. At densities below $\approx 0.5n_0$, the $\chi$EFT band is connected to the Baym-Pethick-Sutherland (BPS) crust EOS \citep{Baym71}.

Using Bayes' theorem, we can write the posterior distributions of the EOS parameters $\bm{\theta}$ and central energy densities $\bm{\varepsilon}$ as 
\begin{equation}
\label{eq:eq1}
p(\bm{\theta}, \bm{\varepsilon} \,|\, \bm{d}, \mathbb{M})
\propto 
p(\bm{\theta} \,|\, \mathbb{M})
~
p(\bm{\varepsilon} \,|\, \bm{\theta}, \mathbb{M})
~
p(\bm{d} \,|\, \bm{\theta}, \mathbb{M}) \,,
\end{equation}
where $\mathbb{M}$ denotes the model including all assumed physics and $\bm{d}$ the dataset used to constrain the EOS, consisting of, e.g., radio, X-ray (NICER), GW, and electromagnetic counterpart (EM) data. Assuming these datasets to be independent of each other, we can separate the likelihoods and write 
\begin{align}
p(\bm{\theta}, \bm{\varepsilon} \,|\, &\bm{d}, \mathbb{M})
\propto 
p(\bm{\theta} \,|\, \mathbb{M})
~
p(\bm{\varepsilon} \,|\, \bm{\theta}, \mathbb{M}) \nonumber\\[1mm]
& \times \prod_{i} p(\Lambda_{1,i}, \Lambda_{2,i}, M_{1,i}, M_{2,i} \,|\, \bm{d}_{\textnormal{GW}, i}, \bm{d}_{\textnormal{EM}, i}) \nonumber\\
& \times \prod_{j} p(M_j, R_j \,|\, \bm{d}_{\textnormal{NICER},j}) \nonumber\\
& \times \prod_{k} p(M_k \,|\, \bm{d}_{\textnormal{radio},k}) \,.
\label{eq:eq2}
\end{align}
Here, $\Lambda_{1,i}$ and $\Lambda_{2,i}$ ($M_{1,i}$ and $M_{2,i}$) are the tidal deformabilities (source-frame component masses) given the GW and EM data $\bm{d}_{\textnormal{GW}, i}$ and $\bm{d}_{\textnormal{EM}, i}$. Furthermore, $\bm{d}_{\textnormal{NICER},j}$ are the mass-radius ($M_j$-$R_j$) NICER data and $\bm{d}_{\textnormal{radio},k}$ the mass data from radio observations. The products run over the number of different observed stars or GW mergers.

In Eq.~(\ref{eq:eq2}), we equate the nuisance-marginalized likelihoods to the nuisance-marginalized posterior distributions from the astrophysical data papers \citep[see][for further discussion of this issue]{Raaijmakers21}. The posterior distributions derived from the X-PSI NICER analysis, which we use in this paper, use a joint uniform prior in mass and radius, if not accounting for the mass prior from radio observations, where available. For a detailed discussion of how the GW parameters are handled, we refer the reader to Sec.~2 of \citet{Raaijmakers21}. In order to speed up convergence, we transform the GW posterior distributions to include the two tidal deformabilities, chirp mass, and mass ratio $q$, fixing the chirp mass to its median value and reweighing such that the LIGO/VIRGO prior distributions on these parameters are uniform. The chirp mass is fixed to its median value because the uncertainties on this parameter are small enough to have no impact on the EOS inference \citep{Raaijmakers21}. By fixing the chirp mass, the central density vector $\bm{\varepsilon}$ will only have one central density per GW event considered, thus the second component's tidal deformability is now a function of  the EOS parameters and the mass ratio, i.e., $\Lambda_{2}(\bm{\theta};q)$.

In \citet{Raaijmakers21}, we included the constraints derived from the radio timing mass measurement of J0740, and from the electromagnetic counterpart of the event GW170817, AT2017gfo. In this work, we do not consider the radio mass measurements separately (they are instead included implicitly as priors on the mass-radius inference with NICER data). We have also chosen not to include the electromagnetic counterpart constraint, given the uncertainties in kilonova modeling \citep[see, e.g.,][]{Raaijmakers21_kn}. With these changes, Eq.~(\ref{eq:eq2}) simplifies to become: 
\begin{align}
p(\bm{\theta}, \bm{\varepsilon} \,|\, &\bm{d}, \mathbb{M})
\propto 
p(\bm{\theta} \,|\, \mathbb{M})
~
p(\bm{\varepsilon} \,|\, \bm{\theta}, \mathbb{M}) \nonumber\\[1mm]
& \times \prod_{i} p(\Lambda_{1,i}, \Lambda_{2,i}, q_i \,|\, \mathcal{M}_c, \bm{d}_{\textnormal{GW}, i}) \nonumber\\
& \times \prod_{l} p_{\textnormal{new}}(M_l, R_l \,|\, \bm{d}_{\textnormal{NICER+radio},l}) \,,
\label{eq:eq3}
\end{align}
where $p_{\textnormal{new}}(M_l, R_l \,|\, \bm{d}_{\textnormal{NICER+radio},l})$ is the redefined NICER likelihood function with the radio observations included in the prior of $\bm{d}_{\mathrm{NICER+radio},l}$ in the cases of J0740 and J0437. With Eq.~(\ref{eq:eq3}) in hand, we sample from the prior distribution $p(\bm{\theta} \,|\, \mathbb{M})\,p(\bm{\varepsilon} \,|\, \bm{\theta}, \mathbb{M})$, compute the corresponding $M$, $R$, and $\Lambda$, and then evaluate the likelihood by applying a kernel density estimation to the posterior distributions from the astrophysical analyses using the nested sampling software \MultiNest\ \citep{Feroz09,Buchner14}. Some modifications are made to the prior distributions of  $p(\bm{\theta} \,|\, \mathbb{M})$ to accommodate the new $\chi$EFT calculations; these are described in more detail in the following section.

\begin{figure}[t!]
    \centering
    \includegraphics[width=0.5\linewidth,clip=]{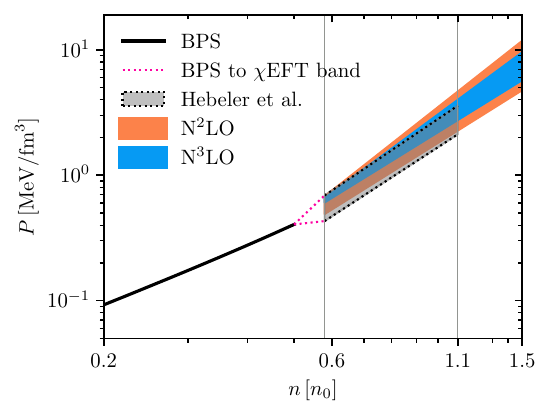}
    \caption{Pressure $P$ as a function of density $n$ for matter in beta equilibrium based on the new $\chi$EFT calculations at N$^2$LO (orange) and N$^3$LO (blue) from \citet{Keller2023} (including contributions from electrons and muons) compared to those from \citet{Hebeler2013} (dotted grey). We also show the BPS crust EOS (solid black line) and the polytropic interpolation of the BPS EOS to the $\chi$EFT band (dotted pink line). The beginning of the $\chi$EFT bands is shown as a gray vertical line at $n = 0.5792 n_0$, and a second vertical line indicates $n = 1.1 n_0$. For simplicity, we only show the BPS EOS to the \citet{Hebeler2013} $\chi$EFT matching, but the transitions to the \citet{Keller2023} \ntwolo\ and \nthreelo\ bands are performed using an identical procedure.}
    \label{fig:chiralEFTbands}
\end{figure}

\subsection{New $\chi$EFT constraints and implementation}
\label{sec:newceft}

In previous works, the $\chi$EFT calculations needed to be combined with an empirical parameterization \citep{Hebeler2013} to go from pure neutron matter to matter in beta equilibrium with a small proton fraction of $\sim 5\%$. The resulting pressure $P$ as a function of density $n$ is shown for the \citet{Hebeler2013} $\chi$EFT band in Fig.~\ref{fig:chiralEFTbands}. As can be seen, the pressure increases to a very good approximation linearly on this log-plot, so that within the $\chi$EFT band it can be described by a single polytrope varying between the minimum and maximum extent of the pressure band.

Figure~\ref{fig:chiralEFTbands} also shows the new $\chi$EFT calculations at \ntwolo\ and \nthreelo\ from \citet{Keller2023}, which are determined directly in beta equilibrium without the need for an empirical parameterization. Since the bands extend to higher densities $n \leq 1.5n_0$, these results also include the small contribution of muons (in addition to electrons) for the pressure of neutron star matter in Fig.~\ref{fig:chiralEFTbands} \citep{Keller2023PhD} (see, e.g., \citealt{Essick2021} for the inclusion of muons). As is evident from Fig.~\ref{fig:chiralEFTbands}, the new \ntwolo\ and \nthreelo\ $\chi$EFT bands can also be effectively parametrized with a single polytrope. This is not surprising, because the density range over which we use $\chi$EFT is small (from $0.5-1.5 n_0$). Moreover, around $n_0$ the density dependence of the pressure is dominated by three-nucleon interactions and in particular the large $c_3$ coupling contribution (see, e.g., \citet{Hebeler2010,Tews2013}). Thus, variations of the $\chi$EFT interactions will give similar density dependencies within the minimum and maximum of the $\chi$EFT band, and because of the limited density range they can again be represented by a single polytrope.

In order to implement the new $\chi$EFT results into the NEoST framework, we therefore fit a single polytrope $P(n) = K \,(n/n_0)^\Gamma$ to the lower and upper pressure limits over the entire density range $0.5 \leq n/n_0 \leq 1.5$. Here $K$ is a constant and $\Gamma$ is the polytropic index. We find that the \ntwolo\ $\chi$EFT band is well reproduced by $K \in [1.814,3.498]$\,MeV\,fm$^{-3}$ and $\Gamma \in [2.391,3.002]$ and the \nthreelo\ $\chi$EFT band by $K \in [2.207,3.056]$\,MeV\,fm$^{-3}$ and $\Gamma \in [2.361,2.814]$. As discussed above, for densities below $n \leq 0.5n_0$, the BPS crust is used, with a log-linear interpolation to the first $\chi$EFT data points at $0.5792 n_0$ for the \citet{Hebeler2013} band. We have checked that this procedure also works for the new $\chi$EFT results, which are also log-linearly interpolated to the BPS crust at $0.5792 n_0$, ensuring that the pressure is never decreasing between the BPS crust and the $\chi$EFT band. This matching between the BPS crust and the $\chi$EFT bands is illustrated in Fig.~\ref{fig:chiralEFTbands}.

\begin{figure}[t!]
    \centering
    \includegraphics[width=0.85\linewidth,clip=]{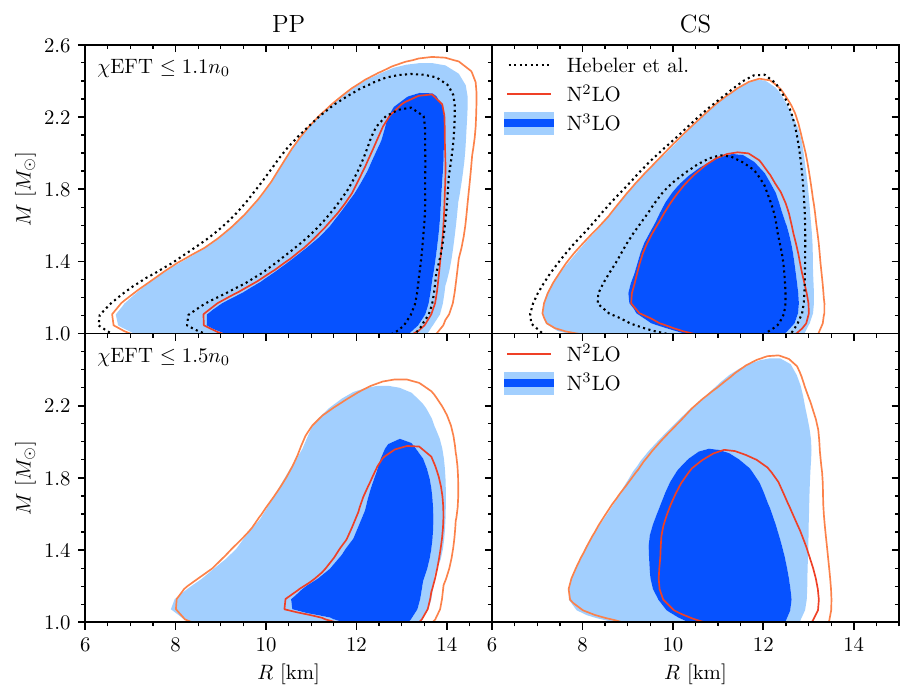}
    \caption{Mass-radius prior distributions for the PP model (left panels) and CS model (right panels). The dark (light) blue region and the inner (outer) curves encompass the 68\% (95\%) credible regions. The top panels compare the priors based on the new $\chi$EFT calculations at N$^2$LO (red) and N$^3$LO (blue) from \citet{Keller2023} to those based on the $\chi$EFT calculations from \citet{Hebeler2013} (dotted black). For the top panels, the $\chi$EFT bands are used up to $1.1 n_0$. In the bottom panels, the prior distributions are shown when using the new $\chi$EFT calculations up to $1.5 n_0$.}
    \label{fig:MR_priors}
\end{figure}

\begin{figure}[t!]
    \centering
    \includegraphics[width=0.85\linewidth,clip=]{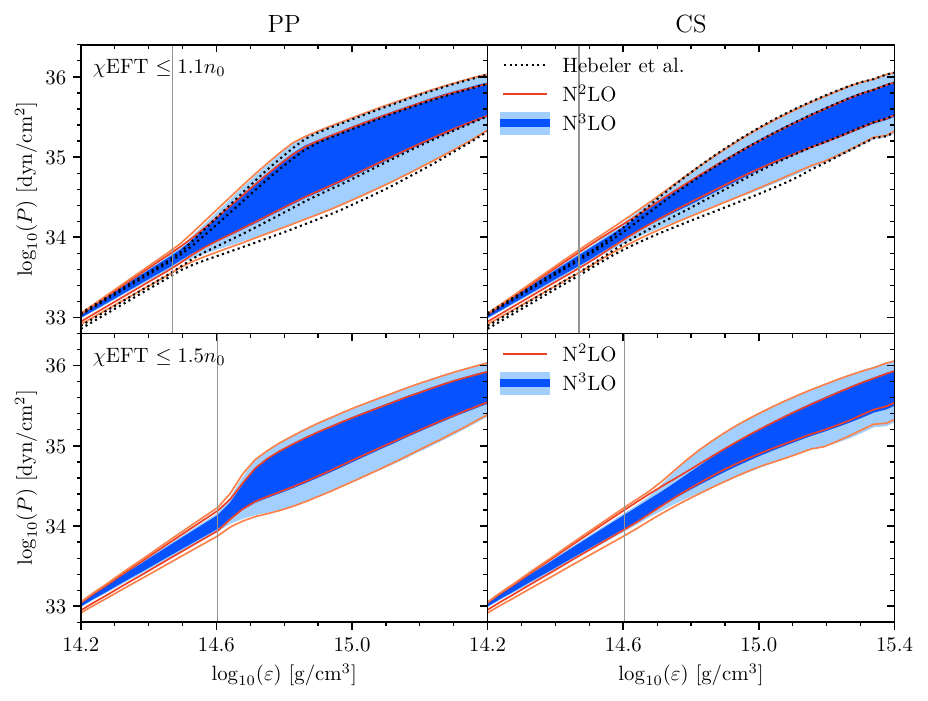}
    \caption{The same as Fig.~\ref{fig:MR_priors}, but for the pressure-energy density prior distributions. The vertical thin lines mark the transition density to the high-density PP and CS extensions.}
    \label{fig:Peps_priors}
\end{figure}

\subsection{New prior distributions}
\label{sec:newpriors}

To generate the PP model, the three polytropic indices are varied within the ranges $\Gamma_1 \in [1,4.5]$, $\Gamma_2 \in [0,8]$, and $\Gamma_3 \in [0.5,8]$, where the first polytrope goes from $1.1 n_0$ to $n_1 \in [1.5,8.3] \, n_0$, the second segment from $n_1$ to $n_2 \in [1.5,8.3] \, n_0$, and the third from $n_2$ to the maximal central density, when the $\chi$EFT band is used up to $1.1 n_0$ \citep[for details see][]{Hebeler2013}. When the $\chi$EFT band is extended to $1.5 n_0$, the parameter ranges are accordingly increased, such that $\Gamma_1 \in [0,8]$, $n_1 \in [2,8.3] \, n_0$, and $n_2 \in [2,8.3] \, n_0$. Note that the first polytropic index was restricted in \citet{Hebeler2013} to limit the EOS variation in the first segment just above the saturation density to reasonable density dependencies. When using $\chi$EFT up to $1.5 n_0$, we remove this limitation. For the EOS and neutron star inference, the pressure as a function of energy density is calculated from the polytropes in number density or mass density using thermodynamic relations \citep[for details see][]{Read2009}.\footnote{We note that as a result of the pressure and number density (or mass density) being the continuous variable in the PP model, the energy density and the chemical potentials are not continuous at the transition between polytropes. However, this has a small effect on bulk properties, such as mass and radius. In \citet{Hebeler2013} we explored this explicitly for the crust to $\chi$EFT matching. Note that the current choice also makes it possible to compare with our previous results. However, this can be improved in the future, e.g., using a generalized piecewise polytropic parameterization \citep{OBoyle2020}.}

The speed of sound parameterization (CS) follows the model detailed in \citet{Greif19}, where $c_s^2 = dP/d\varepsilon$, and  
\begin{equation}
\label{eq:eq4}
    c_{s}^2(x) / c^2=a_1 \mathrm{e}^{-\frac{1}{2}\left(x-a_2\right)^2 / a_3^2}+a_6+\frac{\frac{1}{3}-a_6}{1+\mathrm{e}^{-a_5\left(x-a_4\right)}} \,,
\end{equation}
with $x = \varepsilon/(m_{\mathrm{N}} n_0)$ and the nucleon mass $m_{\mathrm{N}}= 939.565$\,MeV. The parameters $a_1$ to $a_5$ vary within the ranges of $a_1 \in[0.1,1.5]$, $a_2\in[1.5,12]$, $a_3\in[0.075,24]$, $a_4\in[1.5,37]$, $a_5\in[0.1,1]$, and $a_6$ is fixed to continuously match to the $\chi$EFT band polytrope. Further constraints are implemented to guarantee that only EOSs that are causal, $0 \leq c_s^2 \leq c^2$, are included and that the speed of sound approaches the asymptotic value of $c_s^2 = 1/3 c^2$ from below. Moreover, we require the speed of sound up to $1.5 n_0$ to not exceed a limit motivated by Fermi liquid theory \citep[for details see][]{Greif19}:
\begin{equation}
    c_s^2(1.5 \, n_0) / c^2 \leq \frac{1}{m_{\mathrm{N}}^2} \left(3 \pi^2 n \right)^{2/3} \,.
\end{equation}
This is automatically fulfilled for the new $\chi$EFT calculations up to $1.5 n_0$. For the pressure as a function of the energy density needed to solve the Tolman-Oppenheimer-Volkoff (TOV) equations (to obtain mass and radius), we use the same prescription for the polytropes up to the end of the $\chi$EFT band as for the PP model, and then integrate the speed of sound squared $dP/d\varepsilon = c_s^2$ matching to the energy density and with $\varepsilon$ as a continuous variable for higher densities.
Note that for both the PP and CS models we have made the choice to sample uniformly in the space of the EOS parameters. Finally, as previously described in \citet{Hebeler2013,Greif19, Raaijmakers20}, the PP and CS models explicitly allow for first-order phase transitions (with $\Gamma=0$ in PP and a region of $c_s^2 = 0$ in CS).

Implementing the \ntwolo\ and \nthreelo\ $\chi$EFT bands generates the 68\% and 95\% credible region contours of the prior distributions displayed in Fig.~\ref{fig:MR_priors} for mass and radius and Fig.~\ref{fig:Peps_priors} for pressure and energy density. The upper panels in both figures show the case when the new $\chi$EFT bands are used up to $1.1 n_0$, as well as a comparison based on the \citet{Hebeler2013} band (also up to $1.1 n_0$), which was used in our previous EOS inference work \citep{Raaijmakers19,Raaijmakers20,Raaijmakers21}. The lower panels display the prior distributions obtained when using the new $\chi$EFT bands up to $1.5 n_0$. All the priors and the resulting posteriors were calculated for neutron stars with masses $M \geq$~1.0 \Msun: this is theoretically motivated by the description of the early evolution of a neutron star \citep{Strobel_1999} and in agreement with neutron star minimum remnant masses from core-collapse supernova simulations \citep{Janka2008,Fischer2010,Radice2017,Suwa2018}\footnote{While recent spectral modeling of G353.6-0.7 has hinted at the possibility of lower mass neutron stars \citep{Doroshenko22}, this interpretation relies on several critical assumptions (on the distance to the star, which assumes some association with a candidate binary companion, as well as on the spectral modelling of the object and the data sets chosen for the analysis), and is disputed by \citet{Alford2023}.}.

Overall the prior distributions are similar, considering the large range in mass-radius and pressure-energy density. Compared to our previous work using the \citet{Hebeler2013} band, the new \ntwolo\ and \nthreelo\ $\chi$EFT bands shift the mass-radius priors to slightly larger radii. This is consistent with the larger pressures for the new $\chi$EFT bands in Fig.~\ref{fig:chiralEFTbands}. As we increase the transition density to $1.5 n_0$ the prior ranges are narrowed, while keeping similar distributions and mean values for radii. Finally, the prior distributions for the pressure and energy density in Fig.~\ref{fig:Peps_priors} naturally show similar results as for the mass-radius priors, with the tightest credible regions coming from the new $\chi$EFT bands trusted up to $1.5 n_0$ for both the PP and CS parameterizations. 

Since the \citet{Keller2023} \nthreelo\ $\chi$EFT band gives the tightest constraints, with consistent and similar results for the \citet{Hebeler2013} and \citet{Keller2023} \ntwolo\ bands, our analysis from here on will focus on the \nthreelo\ $\chi$EFT band up to the transition densities $1.1 n_0$ and $1.5 n_0$. Results for the \citet{Hebeler2013} and \citet{Keller2023} \ntwolo\ $\chi$EFT bands are provided in the Appendix.

\subsection{Constraints from astrophysical data sets}
\label{sec:astro}

\begin{figure}[t!]
    \centering
    \includegraphics[width=0.85\linewidth,clip=]{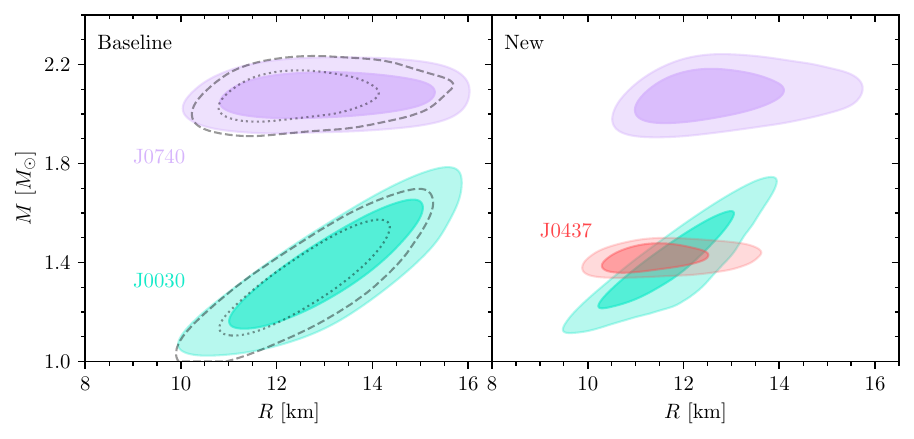}
    \caption{Overview of NICER sources (68\% and 95\% credible regions for mass-radius) for the ``Baseline'' scenario (J0740 results from \citealt{Salmi22} and J0030 NICER-only results from \citealt{Vinciguerra24}) and the ``New'' scenario (J0437 results from \citealt{Choudhury24}, J0740 results from \citealt{Salmi24}, and the \texttt{ST+PDT} solution for J0030, including background constraints, from \citealt{Vinciguerra24}). For the ``Baseline'' scenario, we show for comparison the 68\% (95\%) credible regions from \citet{Riley21} and \citet{Riley19} as dotted (dashed) lines, which were used in \citet{Raaijmakers21}.}
    \label{fig:mrdata}
\end{figure}

In \citet{Raaijmakers21}, we used the following NICER mass-radius and GW mass-tidal deformability posteriors as inputs for our analysis: 
\begin{itemize}
\item The mass and radius for J0030 reported by \citet{Riley19}, using the 2017-2018 NICER data set and the preferred \texttt{ST+PST}\footnote{\texttt{ST-U}, \texttt{ST+PST}, \texttt{ST+PDT}, and \texttt{PDT-U} are different models used in X-PSI that describe the shape and temperature distribution assumed for the hot X-ray emitting spots. For a schematic that illustrates the different models see Fig.~1 of \citet{Vinciguerra23}.} model (68\% CIs $M = 1.34^{+0.15}_{-0.16}\,$\Msun\ and $R = 12.71^{+1.14}_{-1.19}$\,km).  
\item The mass and radius for J0740 reported by \citet{Riley21} from joint modeling of NICER and XMM data, using NICER data from 2018-2020 and the radio-derived mass as a prior, for the preferred \texttt{ST-U} model (68\% CIs $M = 2.08\pm 0.07\,$\Msun\ and $R = 12.39^{+1.30}_{-0.98}$\,km). 
\item{The masses and tidal deformabilities for GW170817 \citep{gw170817} and GW190425 \citep{gw190425}.}
\end{itemize}

To assess the effect of updating our EOS priors and to provide a good baseline for assessing the impact of the new data sets and analyses, we first carry out runs with the older astrophysical inputs, with some small changes. For J0030, we replace the mass and radius posteriors from \citet{Riley19} with the posteriors from the \texttt{ST+PST} NICER-only analysis of the same data set from \citet{Vinciguerra24} (68\% CIs $M = 1.37\pm 0.17\,$\Msun\ and $R = 13.11\pm 1.30$\,km), since these were obtained with an improved analysis pipeline and settings. For J0740, we replace the mass and radius posteriors from \citet{Riley21} with those from \citet{Salmi22}, which treat the background more thoroughly (68\% CIs $M = 2.07\pm 0.07\,$\Msun\ and $R = 12.97^{+1.56}_{-1.39}$\,km). We select the “3C50-3X” case using only NICER data with the 3C50 model \citep{Remillard22} setting a lower limit to the background. Together with the GW mass-tidal deformability posteriors, these form the ``Baseline'' astrophysical scenario in Table~\ref{tab:runplan}.

\begin{figure}[t!]
    \centering
    \includegraphics[width=0.85\linewidth,clip=]{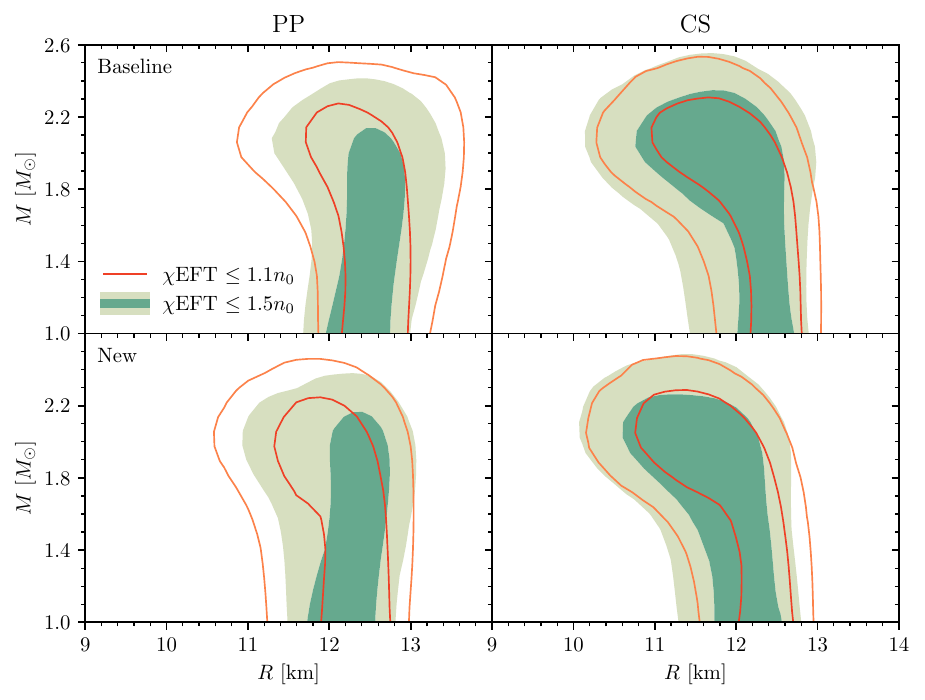}
    \caption{Mass-radius posterior distributions for the ``Baseline'' (upper panels) and ``New'' scenarios (lower panels) using the PP model (left panels) and the CS model (right panels). The dark (light) green regions and the inner (outer) red curves encompass the 68\% (95\%) credible regions using the N$^3$LO $\chi$EFT band up to $1.5 n_0$ and $1.1 n_0$, respectively.}
    \label{fig:MR_baseline}
\end{figure}

We then have several new NICER mass-radius posteriors whose impact we can assess. As in our previous papers, we use results from the X-PSI PPM analysis\footnote{These are derived using a jointly uniform prior distribution on mass and radius (when neglecting the mass prior from radio observations), as discussed in \citet{Riley18,Raaijmakers18}.}. For the high-mass pulsar J0740, we use the inferred mass and radius resulting from joint NICER and XMM analysis, using the 2018-2022 NICER set reported by \citet{Salmi24} (68\% CIs $M = 2.07\pm 0.07\,$\Msun\ and $R = 12.49_{-0.88}^{+1.28}$\,km).  For the 1.4\,\Msun\ pulsar J0437, we use the mass-radius posterior obtained by \citet{Choudhury24} using the 2017-2021 NICER data set, for the preferred \texttt{CST+PDT} model and taking into account limits on the nonsource background (68\% CIs $M = 1.42 \pm 0.04\,$\Msun\ and $R = 11.36^{+0.95}_{-0.63}$\,km).  

For J0030, we consider three alternative sets of mass-radius posteriors from the reanalysis of the 2017-2018 data set reported by \citet{Vinciguerra24}, which supersedes the results of \citet{Riley19}. In addition to the \texttt{ST+PST} NICER-only result used in the ``Baseline'' case, we also consider the two modes preferred in the joint analysis of NICER and XMM data (with XMM being used to place constraints on the nonsource background): \texttt{ST+PDT} (68\% CIs $M = 1.40^{+0.13}_{-0.12}\,$\Msun\ and $R = 11.71^{+0.88}_{-0.83}$\,km) and \texttt{PDT-U} (68\% CIs $M = 1.70^{+0.18}_{-0.19}\,$\Msun\ and $R = 14.44^{+0.88}_{-1.05}$\,km). \texttt{PDT-U} is preferred by the Bayesian evidence, although \citet{Vinciguerra24} caution that higher resolution runs are required to check the robustness of the joint NICER-XMM runs. However, the \texttt{ST+PDT} results are more consistent with the magnetic field geometry inferred for the gamma-ray emission for this source (\citealt{Kalapotharakos21}, as discussed in \citealt{Vinciguerra24}) and the inferred mass and radius for this mode are most consistent with the new results for J0437. For these reasons, we deem this at present---with all reserve and pending further analysis---to be the most likely solution for J0030.

Therefore, the combination of posteriors from \texttt{ST+PDT} for J0030, with \citet{Salmi24} for J0740 and \citet{Choudhury24} for J0437, as well as the GW data sets \citep{gw170817,gw190425} forms the ``New'' astrophysical scenario. The impact of the other J0030 solutions is investigated in the Appendix.
The GW results are unchanged compared to the ``Baseline'' scenario since there are as yet no new mass-tidal deformability results to be included. Figure~\ref{fig:mrdata} shows the different mass-radius posteriors used in the ``Baseline'' and ``New'' scenarios.

\begin{figure}[t!]
    \centering
    \includegraphics[width=0.85\linewidth,clip=]{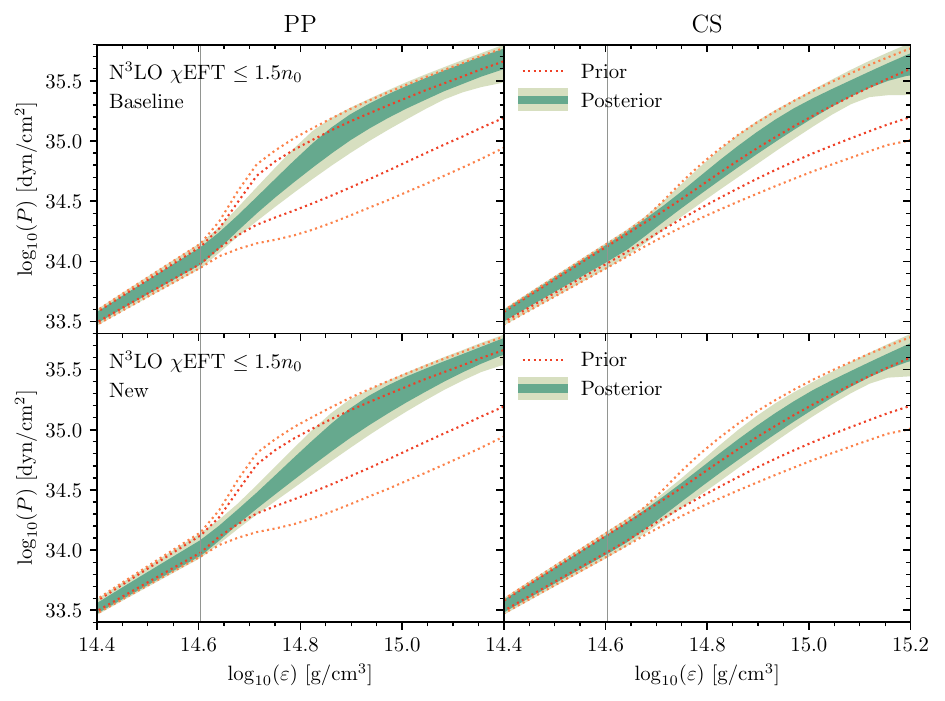}
    \caption{Pressure-energy density posterior distributions for the ``Baseline'' (upper panels) and ``New'' scenario (lower panels) using the PP model (left panels) and the CS model (right panels). The dark (light) green regions encompass the 68\% (95\%) credible regions, while the dotted red lines represent the corresponding prior distributions using the N$^3$LO $\chi$EFT band up to $1.5 n_0$. The vertical thin lines mark the transition density to the high-density PP and CS extensions.}
    \label{fig:Peps_baseline}
\end{figure}

\section{Results: Impact of new prior distributions and new astrophysical constraints}
\label{sec:results}

In this section, we investigate the combined impact of the new NICER results and the new \nthreelo\ $\chi$EFT results. We first analyze the posterior inferences on the ``Baseline'' scenario to understand how constraining the \nthreelo\ $\chi$EFT calculations are up to a given density. We first analyze the posterior inferences on the ``Baseline'' scenario to understand the effects of the different densities up to which the \nthreelo\ $\chi$EFT calculation is trusted. We then compare the inferences on the ``New'' to the ``Baseline'' scenario, to study the impact of new astrophysical constraints on the inferred dense matter EOS and neutron star properties. The results for the other $\chi$EFT bands and for the sensitivities to the J0030 results are given in the Appendix.

In Figs.~\ref{fig:MR_baseline} and \ref{fig:Peps_baseline}, we show the mass-radius and pressure-energy density posteriors for the ``Baseline'' and ``New'' scenarios, comparing the new $\chi$EFT bands at \nthreelo\ with transition densities $1.1  n_0$ and $1.5 n_0$. For the CS model, Fig.~\ref{fig:MR_baseline} shows that trusting the $\chi$EFT results up to $1.1 n_0$ and $1.5 n_0$ predicts similar mass-radius confidence regions with the $1.5 n_0$ results tending to smaller radii (especially for masses below 1.6 \Msun) for both the ``Baseline'' and ``New'' scenarios. For the PP model, however, trusting the \nthreelo\ $\chi$EFT up to $1.5 n_0$ shrinks the radius posteriors compared to $1.1 n_0$. Moreover, the mass-radius posteriors predict lower maximum masses when $\chi$EFT is trusted up to $1.5 n_0$ compared to $1.1 n_0$ at the 68\% and 95\% CIs. Compared to the prior distributions, which include a broad range of softer EOSs with smaller radii, the pressure-energy density posterior distributions in Fig.~\ref{fig:Peps_baseline} prefer stiffer EOSs to support heavy-mass pulsars and radii around 12.5\,km.

We next compare the mass-radius posterior distributions for the ``New'' scenario to the ``Baseline'' scenario, leaving the exploration of the effect of the different J0030 results to the Appendix. For the ``New'' scenario, the mass-radius posterior regions for $1.1 n_0$ and $1.5 n_0$ in Fig.~\ref{fig:MR_baseline} are shifted and/or narrowed to smaller radii compared to the ``Baseline'' scenario for both the PP and CS models. This shifting/narrowing is due to the addition of the J0437 NICER results as well as to the fact that the J0030 \texttt{ST+PDT} results are consistent with the inferred radius of J0437.

\begin{figure}[t!]
    \centering
    \includegraphics[width=0.85\linewidth,clip=]{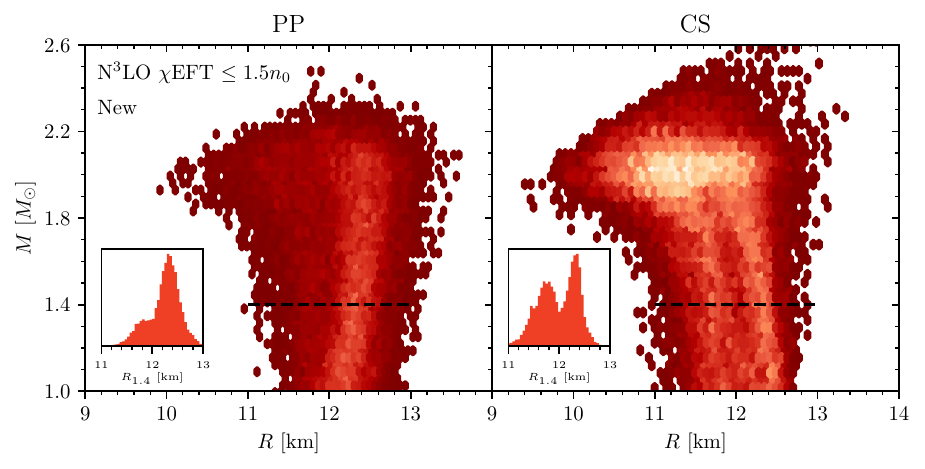}
    \caption{Mass-radius posterior distributions showing the 2D histogram for the ``New'' scenario using the N$^3$LO $\chi$EFT band up to $1.5 n_0$. The dark shaded hexagons indicate a lower number of mass-radius samples, while the lighter shaded hexagons indicate a higher number of mass-radius samples. The insets show the radius distribution for a 1.4 \Msun\ star, indicated by the black dashed lines.}
    \label{fig:MR_Newcomp_N3LO_1.5}
\end{figure}

Interestingly, if the \nthreelo\ $1.5 n_0$ posterior samples of the ``New'' scenario are cast as a 2D histogram shown in  Fig.~\ref{fig:MR_Newcomp_N3LO_1.5}, the posteriors show hints of a bimodal-like distribution when the new NICER results are folded in. Figure~\ref{fig:MR_Newcomp_N3LO_1.5} shows a bimodal-like structure centered around 12\,km in the CS model. However, for the PP model, the bimodal-like structure is less pronounced. 

The bimodal-like distribution also manifests in the posterior distributions of the parameter, $K$, which is the constant that matches to the $\chi$EFT pressure. Additionally, this structure appears in the pressure posteriors at the intermediate densities $2 n_0$ and $3 n_0$, plotted in Fig.~\ref{fig:Press_New2}. Here, it is clearly present for the \nthreelo\ $\chi$EFT $\leq 1.5 n_0$ band for the CS models at $2 n_0$, but only hinted at in the PP model at $3 n_0$. Despite the bimodal-like structure also being present in the mass-radius posteriors for the \nthreelo\ $\chi$EFT $\leq 1.1 n_0$, this effect is not seen in the histogram plots in Fig.~\ref{fig:Press_New2}. Additionally, we also find that the bimodal-like distribution is more strongly present with the \ntwolo\ $\chi$EFT band, see the Appendix for further details. This bimodal-like distribution suggests a tension between the posteriors of J0740, which favors higher radii, and the GW results in combination with J0437, which favor lower radii posteriors. We also find that this tension is enhanced by the J0030 \texttt{ST+PDT} mass-radius posteriors due to the strong overlap with J0437.

\section{Discussion and conclusions}
\label{sec:discuss}

In this Letter, we have investigated the constraints on the EOS posed by new joint mass–radius estimates from NICER analysis for J0437 \citep{Choudhury24}, J0740 \citep{Salmi24} and J0030 \citep{Vinciguerra24}, combined with tidal deformabilities measured during binary neutron star mergers with gravitational waves, and using new $\chi$EFT calculations \citep{Keller2023} up to $1.5 n_0$. In Table \ref{tab:ranges} we summarize the results, including the constraints on the radius of a 1.4 and 2.0\,\Msun\ neutron star, the maximum mass and radius of a nonrotating neutron star $M_\mathrm{TOV}$, as well as the central energy density, central density, and pressure for these masses. 

\subsection{Implications for the dense matter EOS}

\begin{figure}[t!]
    \centering
    \includegraphics[width=0.8\linewidth,clip=]{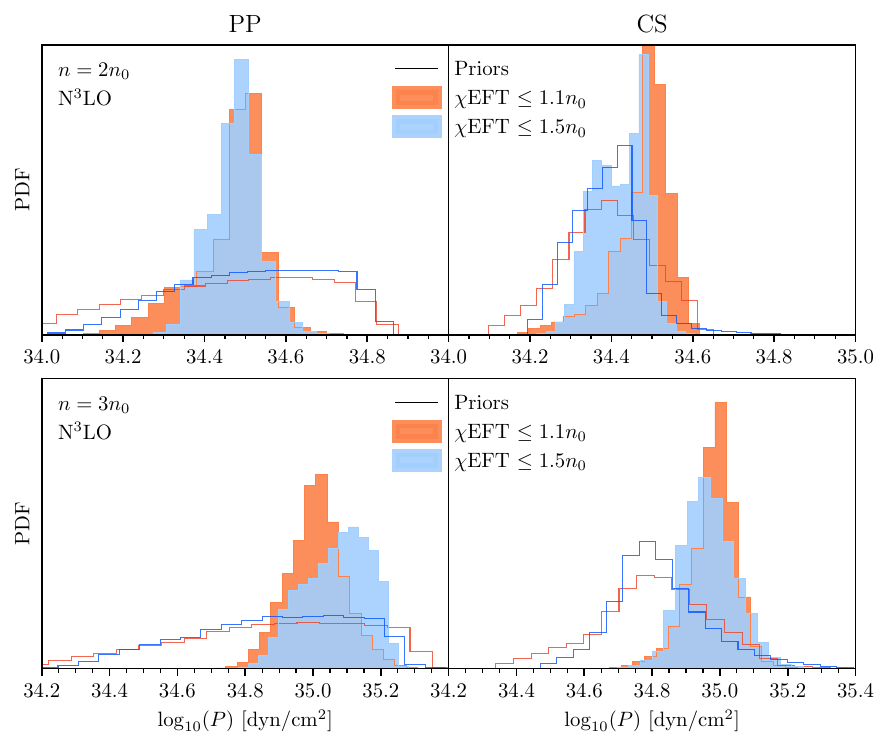}
    \caption{Prior distributions (lines) and posterior distributions (colored regions) for the pressure at $2n_0$ (upper panels) and $3n_0$ (lower panels). Results are shown for the PP model (left panels) and the CS model (right panels) using the $\chi$EFT bands at N$^3$LO up to $1.1 n_0$ (orange) and $1.5 n_0$ (light blue). The posterior distributions are for the ``New'' scenario.}
    \label{fig:Press_New2}
\end{figure}

As in \citet{Raaijmakers21}, we study the changes from the prior to posterior distributions for the pressure at $2 n_0$ and $3 n_0$ in Fig.~\ref{fig:Press_New2}. At densities just above the $\chi$EFT bands, this makes the impact of the astrophysical observations for constraining the dense matter EOS particularly visible. For the PP model at $2 n_0$, in Fig.~\ref{fig:Press_New2}, the posterior for the pressure is in the central range of the prior, but substantially narrower. In contrast, the CS model prior is already narrower, and the posterior is over a similar range, centered around $10^{34.4}$\,dyn/cm$^2$. This demonstrates the overall consistency among the pressure posteriors for both the PP and CS models. Moreover, as expected, we find a narrower pressure posterior when trusting $\chi$EFT up to $1.5 n_0$, especially for the PP model. The pressure posteriors at $3 n_0$ are significantly narrowed compared to the broad priors, where the astrophysical results clearly prefer stiffer EOSs at the upper range of the priors in all cases. Additionally, at $3 n_0$, the pressures for all $\chi$EFT assumptions consistently predict values centered around $10^{35}$\,dyn/cm$^2$. Finally, the presence of a bimodal-like structure in some of the pressure posteriors of the \nthreelo\ $\chi$EFT bands up to $1.5 n_0$ (see Sec.~\ref{sec:results}) suggests that the inferred neutron star EOS could be described equally well by both a relatively softer or stiffer EOS between $2-3n_0$. 

\begin{figure}[t!]
    \centering
    \includegraphics[width=0.85\linewidth,clip=]{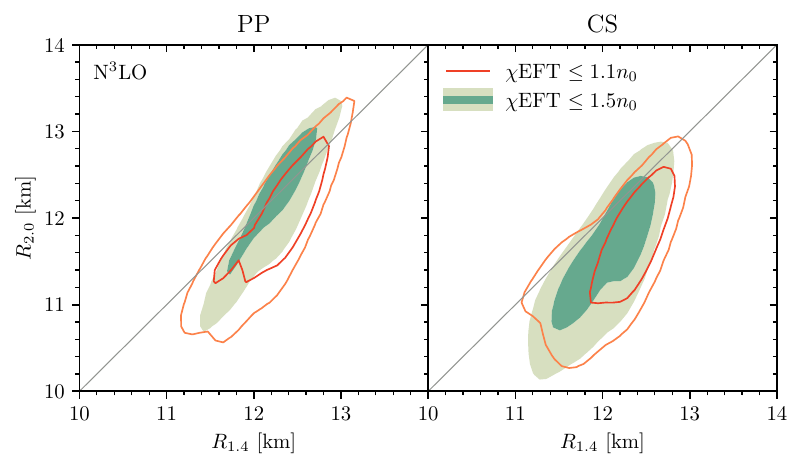}
    \caption{Correlation between the radius $R_{2.0}$ and $R_{1.4}$ of a $2.0\,$\Msun\ and $1.4\,$\Msun\ neutron star, respectively, for the ``New'' scenario. Results are shown for the PP model (left panel) and the CS model (right panel) using the $\chi$EFT bands at N$^3$LO up to $1.1 n_0$ (red) and $1.5 n_0$ (green).}
    \label{fig:R2.0_vs_R1.4}
\end{figure}

An interesting quantity explored in \citet{Drischler2021} is the difference $\Delta R = R_{2.0} - R_{1.4}$ between the radius $R_{2.0}$ and $R_{1.4}$ of a $2.0\,$\Msun\ and $1.4\,$\Msun\ neutron star respectively. In particular, \citet{Drischler2021} pointed out that the sign of $\Delta R$ could be an indicator that the underlying EOS softens (if negative) or stiffens (if positive) at high densities. Our results for $\Delta R$ are given in Table~\ref{tab:ranges} and explored with the correlation plot in Fig.~\ref{fig:R2.0_vs_R1.4} for the posteriors of our ``New'' scenario. We find that for the PP model the preference for $\Delta R$ being positive or negative (albeit with large uncertainties) depends on the transition density, but for the CS model, negative $\Delta R$ is preferred for all transition densities. Note that \citet{Choudhury24} reported $\Delta R = 1.13^{+1.59}_{-1.08}$\,km (68\% CI). At first glance our inferred $\Delta R$ values and the $\Delta R$ obtained by \citet{Choudhury24} appear to be in tension with one another, as our results are centered around $\Delta R = -0.4 $ to $-0.6$\,km (CS model) or slightly positive or negative $\Delta R$ (PP model). However, the uncertainties are larger and the value quoted in \citet{Choudhury24} was computed directly from the radius credible intervals for J0437 and the radius inferred for J0740 by \citet{Salmi24} (subtraction of the median values and simple compounding of the uncertainties), independent of any EOS model. Our results here are derived from the posteriors of the EOS models and their respective priors in addition to the full posteriors for J0030, J0437, J0740, GW170817, and GW190425. These all jointly push the inferred $\Delta R$ to smaller central values than the value quoted in \citet{Choudhury24}.

\subsection{Implications for neutron star maximum mass}\label{sec:max_mass}

A key quantity relating to the dense matter EOS is the maximum mass of a nonrotating neutron star, $M_{\rm TOV}$. This defines the boundary between neutron stars and black holes, and is necessary to our understanding of stellar evolution, supernovae, and compact object mergers. 

\begin{figure}[t!]
    \centering
    \includegraphics[width=0.85\linewidth,clip=]{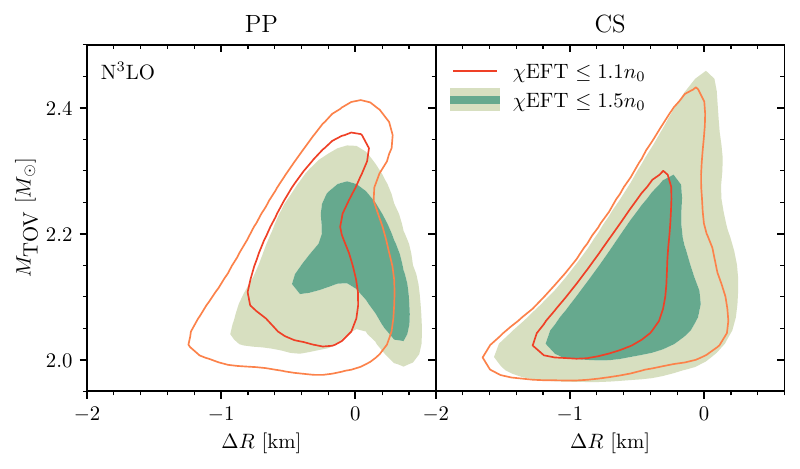}
    \caption{The same as Fig.~\ref{fig:R2.0_vs_R1.4} but for the maximum mass $M_\text{TOV}$ vs. $\Delta R = R_{2.0} - R_{1.4}$.}
    \label{fig:deltaR_maxM_new2}
\end{figure}

In Fig.~\ref{fig:deltaR_maxM_new2} we give the joint posterior distribution of $M_{\rm TOV}$ and $\Delta R$ for the ``New'' scenario using the \nthreelo\ $\chi$EFT band up to $1.1 n_0$ and $1.5 n_0$. This shows that the maximum mass is predicted to be below around 2.4\,\Msun\ (95\% CI) for all $\chi$EFT assumptions, with slightly larger values for the CS models. Moreover, there is a general trend of increasing $M_\text{TOV}$ with larger values of $\Delta R$. In addition, in Fig.~\ref{fig:deltaR_maxM_new2_appendix} of the Appendix, we observe that the bounds on $M_\text{TOV}-\Delta R$ are mostly unaffected when going from \ntwolo\ to \nthreelo\, and that the bounds on $M_\text{TOV}$ are very similar for the PP and CS models for both \ntwolo\ and \nthreelo. 

In order to assess the impact of the new astrophysical results on $M_\mathrm{TOV}$, we show in Table~\ref{tab:ranges} the 95\% CIs inferred using the \nthreelo\ $\chi$EFT band for both transition densities---$1.1 n_0$ and $1.5 n_0$---and both the ``Baseline'' and ``New'' astrophysical scenarios. For the PP model, we find that the 95\% CI on $M_\mathrm{TOV}$ is constrained to $2.17^{+0.15}_{-0.17}$\,\Msun\ and $2.15^{+0.14}_{-0.16}$\,\Msun\ for the ``Baseline'' and ``New'' scenarios, respectively. For the CS model, the ``New'' $M_\mathrm{TOV}$ posteriors are again shifted to lower masses when compared to the ``Baseline'' scenario. In particular, we find that ``New'' constrains the 95\% CI on $M_\mathrm{TOV}$ to $2.08^{+0.28}_{-0.16}$\,\Msun, while  the ``Baseline'' scenario constrains $M_{\mathrm{TOV}}$ to $2.11^{+0.31}_{-0.16}$\,\Msun. However, despite ``New'' tending to lower maximum masses, all of the $M_\mathrm{TOV}$ posteriors strongly overlap, suggesting that the impact of the new astrophysical results on $M_\mathrm{TOV}$ is small and that it remains strongly dependent on the mass-radius measurements of high mass pulsars like J0740.

\begin{table*}[t!]
\caption{Key quantities for the posterior distributions for the different astrophysical scenarios using the N$^3$LO $\chi$EFT bands up to $1.1 n_0$ and $1.5 n_0$: the radius of a $1.4$\,\Msun\ and a $2$\,\Msun\ neutron star, $\Delta R = R_{2.0} - R_{1.4}$, the maximum mass of a nonrotating neutron star $M_{\rm TOV}$, and the corresponding radius to the maximum mass of a nonrotating neutron star $R_{\rm TOV}$. We also list the inferred central energy densities $\varepsilon_c$, central densities $n_c/n_0$, and the corresponding central pressures $P_c$ of $M_{\rm TOV}$, a $1.4$\,\Msun, and a $2$\,\Msun\ neutron star. Radii are given in\,km, $M_{\rm TOV}$ in \Msun, and the central energy densities and pressures in g/cm$^3$ and dyn/cm$^{2}$, respectively. The upper and lower values correspond to the $95\%$ CI.}
\centering
\begin{tabular}{|c|cc|cc|}
\hline
& \multicolumn{2}{c|}{N$^3$LO $\chi$EFT $\le 1.1n_0$} & \multicolumn{2}{c|}{N$^3$LO $\chi$EFT $\le 1.5n_0$} \\ 
& Baseline & New & Baseline & New  \\
\hline
& \multicolumn{4}{c|}{PP model} \\ 
\hline
$R_{1.4}$ & $12.58^{+0.67}_{-0.75}$ & $12.30^{+0.55}_{-1.04}$ & $12.48^{+0.57}_{-0.67}$ & $12.28^{+0.50}_{-0.76}$\\
$R_{2.0}$ & $12.40^{+1.12}_{-1.26}$ & $11.99^{+0.93}_{-1.17}$& $12.59^{+0.78}_{-1.24}$ & $12.33^{+0.70}_{-1.34}$\\
$\Delta R$ & $-0.16^{+0.48}_{-0.76}$ & $-0.28^{+0.47}_{-0.71}$& $0.14^{+0.24}_{-0.77}$ & $0.05^{+0.29}_{-0.78}$\\
\hline
$M_{\rm TOV}$ & $2.27^{+0.16}_{-0.26}$ & $2.15^{+0.20}_{-0.20}$ & $2.17^{+0.15}_{-0.17}$ & $2.15^{+0.14}_{-0.16}$ \\
$R_{\rm TOV}$ & $12.03^{+1.51}_{-1.47}$& $11.58^{+1.21}_{-1.21}$& $12.55^{+0.89}_{-1.68}$& $12.22^{+0.83}_{-1.62}$\\
\hline
$\log_{10}$($\varepsilon_{c,\rm{TOV}})$ & $15.13^{+0.26}_{-0.23}$& $15.17^{+0.24}_{-0.20}$& $14.99^{+0.34}_{-0.14}$& $15.04^{+0.31}_{-0.15}$\\
$n_{c,\rm{TOV}}/n_0$ & $4.18^{+2.51}_{-1.50}$ & $4.52^{+2.40}_{-1.45}$ & $3.24^{+2.85}_{-0.79}$ & $3.58^{+2.73}_{-0.90}$\\
$\log_{10}$($P_{c,\rm{TOV}})$ & $35.66^{+0.36}_{-0.35}$& $35.70^{+0.33}_{-0.33}$& $35.44^{+0.48}_{-0.26}$& $35.52^{+0.42}_{-0.26}$\\
\hline
$\log_{10}$($\varepsilon_{c,1.4})$ & $14.87^{+0.11}_{-0.11}$& $14.91^{+0.11}_{-0.09}$& $14.85^{+0.11}_{-0.08}$ & $14.87^{+0.13}_{-0.07}$\\
$n_{c,1.4}/n_0$ & $2.57^{+0.66}_{-0.54}$ & $2.80^{+0.78}_{-0.48}$ & $2.47^{+0.68}_{-0.38}$  & $2.62^{+0.81}_{-0.37}$\\
$\log_{10}$($P_{c,1.4})$ & $34.96^{+0.15}_{-0.14}$& $35.02^{+0.18}_{-0.11}$& $34.96^{+0.14}_{-0.10}$& $34.99^{+0.17}_{-0.09}$\\
\hline
$\log_{10}$($\varepsilon_{c,2.0})$ & $15.02^{+0.21}_{-0.17}$& $15.07^{+0.19}_{-0.15}$& $14.95^{+0.22}_{-0.11}$&$14.99^{+0.23}_{-0.11}$\\
$n_{c,2.0}/n_0$ & $3.43^{+1.75}_{-1.01}$ & $3.85^{+1.69}_{-1.0}$ & $3.01^{+1.66}_{-0.59}$  & $3.27^{+1.86}_{-0.64}$\\
$\log_{10}$($P_{c,2.0})$ & $35.39^{+0.32}_{-0.25}$ & $35.49^{+0.30}_{-0.22}$& $35.31^{+0.32}_{-0.15}$ & $35.37^{+0.35}_{-0.15}$\\
\hline
& \multicolumn{4}{c|}{CS model} \\ 
\hline
$R_{1.4}$ & $12.44^{+0.41}_{-0.9}$ & $12.29^{+0.47}_{-1.03}$& $12.29^{+0.42}_{-0.94}$ & $12.01^{+0.56}_{-0.75}$\\
$R_{2.0}$ & $11.91^{+0.8}_{-1.25}$& $11.69^{+0.84}_{-1.12}$& $11.87^{+0.89}_{-1.35}$& $11.55^{+0.94}_{-1.09}$\\
$\Delta R$ & $-0.52^{+0.52}_{-0.76}$ & $-0.58^{+0.61}_{-0.73}$& $-0.40^{+0.60}_{-0.82}$ & $-0.46^{+0.59}_{-0.76}$\\
\hline
$M_{\rm TOV}$ & $2.11^{+0.28}_{-0.16}$ & $2.08^{+0.25}_{-0.17}$ & $2.11^{+0.31}_{-0.16}$ & $2.08^{+0.28}_{-0.16}$\\ 
$R_{\rm TOV}$ &$11.16^{+1.18}_{-1.15}$& $10.97^{+1.17}_{-1.02}$& $11.25^{+1.38}_{-1.32}$&$10.94^{+1.37}_{-1.04}$\\
\hline
$\log_{10}$($\varepsilon_{c,\rm{TOV}})$ & $15.38^{+0.09}_{-0.09}$& $15.38^{+0.09}_{-0.09}$& $15.38^{+0.09}_{-0.14}$& $15.38^{+0.09}_{-0.09}$\\
$n_{c,\rm{TOV}}/n_0$ & $6.53^{+1.41}_{-1.11}$ & $6.64^{+1.35}_{-1.04}$ & $6.65^{+1.51}_{-1.45}$ & $6.82^{+1.44}_{-1.30}$\\
$\log_{10}$($P_{c,\rm{TOV}})$ & $35.89^{+0.24}_{-0.40}$ & $35.90^{+0.24}_{-0.37}$& $35.85^{+0.30}_{-0.46}$& $35.88^{+0.27}_{-0.42}$\\
\hline
$\log_{10}$($\varepsilon_{c,1.4})$ & $14.90^{+0.12}_{-0.07}$& $14.92^{+0.11}_{-0.07}$& $14.91^{+0.12}_{-0.08}$& $14.94^{+0.10}_{-0.09}$\\
$n_{c,1.4}/n_0$ & $2.76^{+0.78}_{-0.38}$ & $2.88^{+0.79}_{-0.41}$ & $2.83^{+0.83}_{-0.46}$ & $3.02^{+0.69}_{-0.51}$\\
$\log_{10}$($P_{c,1.4})$ & $35.0^{+0.18}_{-0.09}$& $35.03^{+0.18}_{-0.10}$& $35.02^{+0.18}_{-0.10}$& $35.07^{+0.15}_{-0.11}$\\
\hline
$\log_{10}$($\varepsilon_{c,2.0})$ & $15.13^{+0.19}_{-0.14}$& $15.16^{+0.18}_{-0.15}$& $15.12^{+0.21}_{-0.17}$& $15.16^{+0.19}_{-0.17}$\\
$n_{c,2.0}/n_0$ & $4.26^{+1.87}_{-1.06}$ & $4.49^{+1.79}_{-1.11}$ & $4.19^{+2.07}_{-1.18}$ & $4.51^{+1.90}_{-1.26}$\\
$\log_{10}$($P_{c,2.0})$ &$35.54^{+0.33}_{-0.21}$& $35.58^{+0.30}_{-0.21}$& $35.53^{+0.36}_{-0.24}$& $35.59^{+0.31}_{-0.25}$\\
\hline
\end{tabular}
\label{tab:ranges}
\end{table*}

The largest systematic uncertainty in the astrophysical constraints relates to the uncertainty in the inferred mass-radius for J0030, which is now known to have multiple geometric modes with different associated masses and radii. In our analysis, we have considered two different modes identified in \citet{Vinciguerra24}: one for which background constraints are not taken into account (\texttt{ST+PST}) in the ``Baseline'' scenario, and one mode that emerges when background constraints are applied using joint analysis of NICER and XMM data (\texttt{ST+PDT}) in the ``New'' scenario. The results in the Appendix, considering other scenarios for J0030, show the sensitivity of the inferred mass-radius to the J0030 results and thus the importance of background constraints. Background constraints are important because by putting bounds on the amount of unpulsed emission coming from other sources, they can rule out certain combinations of hot spot geometry and compactness, (see, e.g., \citealt{Salmi22}). We cautiously favor the \texttt{ST+PDT} solution over the \texttt{PDT-U} one due to its consistency with both the J0437 mass-radius results and multi-wavelength pulsar emission models. However, as discussed in \citet{Vinciguerra24}, higher resolution runs are needed to confirm the robustness of the joint NICER and XMM analysis. Given the anticipated high computational cost, these runs have been deferred, awaiting the preparation of a new larger data set covering all currently available NICER data (\citealt{Vinciguerra24} used a data set consisting of NICER data from 2017-2018). Nonetheless, the importance of these posteriors to the dense matter EOS analysis is evident.

\subsection{Systematic uncertainties}

Our results are conditional on the choices made for the EOS models and the astrophysical constraints analyzed. The sensitivity of our inferences on the choice of nuclear physics priors is studied by considering the new $\chi$EFT calculations of \citet{Keller2023} at \nthreelo\ (with \ntwolo\ studied in the Appendix), trusted up to $1.1 n_0$ and $1.5 n_0$. As expected, we find that the \nthreelo\ bands result in tighter posterior constraints when we trust them up to $1.5 n_0$. Interestingly, a bimodal-like distribution manifests in the EOS posteriors for both the PP and CS models (see Sec.~\ref{sec:results}). Although our inferences suggest that this bimodal-like structure is due to a tension between the posteriors of J0740 and those of J0437 and GW170817, further investigations to better understand its origin are needed, which we leave for future work.

In addition to exploring the sensitivities of the EOS models to the new $\chi$EFT bands, we considered the two different high-density PP and CS extensions. From Table~\ref{tab:ranges}, we find that the CS model consistently predicts lower radii for $R_{1.4}$, $R_{2.0}$, and $R_\mathrm{TOV}$ than the PP model. Therefore, relative to the PP model, the CS model prefers softer neutron star EOSs, in agreement with \citet{Raaijmakers21}. Table~\ref{tab:ranges} additionally reveals that $M_\mathrm{TOV}$, $R_\mathrm{TOV}$, and the corresponding central energy densities, central densities, and pressures are noticeably sensitive to increasing $\chi$EFT transition densities in the PP model, but exhibit a much lower sensitivity to the higher transition density in the CS model.

\subsection{Summary and future prospects}

In this Letter we have studied the impact of the new NICER data and analysis, especially for J0437 and J0740, as well as the choice of J0030 mass-radius posteriors, on the inferred neutron star EOS. Our work shows that the new $\chi$EFT results, especially the extension of the \nthreelo\ band to $1.5 n_0$, tighten the EOS posteriors significantly. In Table~\ref{tab:ranges}, we summarize our results of the key quantities for the posterior distributions for the different astrophysical scenarios. In particular, we find the radius of a 1.4\,\Msun\ (2.0\,\Msun) neutron star is constrained to the 95\% credible ranges $12.28^{+0.50}_{-0.76}\,$km ($12.33^{+0.70}_{-1.34}\,$km) for the PP model and $12.01^{+0.56}_{-0.75}\,$km ($11.55^{+0.94}_{-1.09}\,$km) for the CS model, for what we consider our most likely ``New'' scenario. In this scenario, the maximum mass of neutron stars is predicted to be $2.15^{+0.14}_{-0.16}\,$\Msun\ and $2.08^{+0.28}_{-0.16}\,$\Msun\ for the PP and CS models, respectively. 

NICER continues to collect more data on all of its targets, including four sources---some of which also have mass priors from radio pulsar timing---for which mass-radius inferences have yet to be published. As we obtain data from more sources and tighter mass-radius inferences, the EOS constraints will continue to improve. New heavy-mass pulsar measurements from radio timing and new GW measurements of tidal deformability are also anticipated, with the LIGO-VIRGO-KAGRA collaboration's next observing run (O4b) starting in April 2024.

\section*{Acknowledgments}

This work was supported in part by NASA through the NICER mission and the Astrophysics Explorers Program.  The work of C.P.W. and N.R. was supported by NASA Grant No.~80NSSC22K0092. The work of K.H., J.K., M.M., A.S. and I.S. was supported by the European Research Council (ERC) under the European Union's Horizon 2020 research and innovation programme (Grant Agreement No.~101020842) and by the Deutsche Forschungsgemeinschaft (DFG, German Research Foundation)---Project-ID 279384907---SFB 1245. A.L.W., D.C., T.S. and S.V. acknowledge support from ERC Consolidator Grant No.~865768 AEONS (PI: Watts). Computational work was carried out on the HELIOS cluster including dedicated nodes funded via this ERC CoG as well as on the high-performance computing cluster Lichtenberg at the NHR Centers NHR4CES at TU Darmstadt. S.G. acknowledges the support of the CNES. J.M.L. acknowledges support from the US Department of Energy under Grant DE-FG02-87ER40317. We acknowledge extensive use of NASA’s Astrophysics Data System (ADS) Bibliographic Services and the ArXiv.

\begin{table}[h!]
   \centering
   \caption{Overview of the astrophysical constraints used in the baseline and the three new scenarios (for details see text), the abbreviation `w bkg' means `with background constraints'. The ``Baseline'' and ``New'' scenarios are discussed in the main text. The ``New 2'' and ``New 3'' scenarios include different posteriors for J0030 from \citet{Vinciguerra24}.}
   \label{tab:runplan}
   \begin{tabular}{|l|c|c|c|c|c|c|c|}
   \hline
        & GW & \multicolumn{2}{c|}{J0740} & \multicolumn{3}{c|}{J0030} & J0437 \\
        \hline
        & GW170817 & \multicolumn{2}{c|}{\texttt{ST-U}} & \multicolumn{3}{c|}{\citet{Vinciguerra24}} & \citet{Choudhury24} \\
        & + & NICER w bkg & NICER $\times$ XMM & NICER & \multicolumn{2}{c|}{NICER $\times$ XMM} & NICER w bkg \\
        & GW190425 & \citet{Salmi22} & \citet{Salmi24} & \texttt{ST+PST} & \texttt{ST+PDT} & \texttt{PDT-U} & \texttt{CST+PDT} \\
        \hline
        Baseline & $\times$ & $\times$ & & $\times$ & & & \\
        New & $\times$ & & $\times$ & & $\times$ & & $\times$ \\
        \hline
        New 2 & $\times$ & & $\times$ & $\times$ & & & $\times$ \\
        New 3 & $\times$ & & $\times$ & & & $\times$ & $\times$ \\
       \hline
   \end{tabular}
\end{table}

\begin{figure}[h!]
    \centering
    \includegraphics[width=0.85\linewidth,clip=]{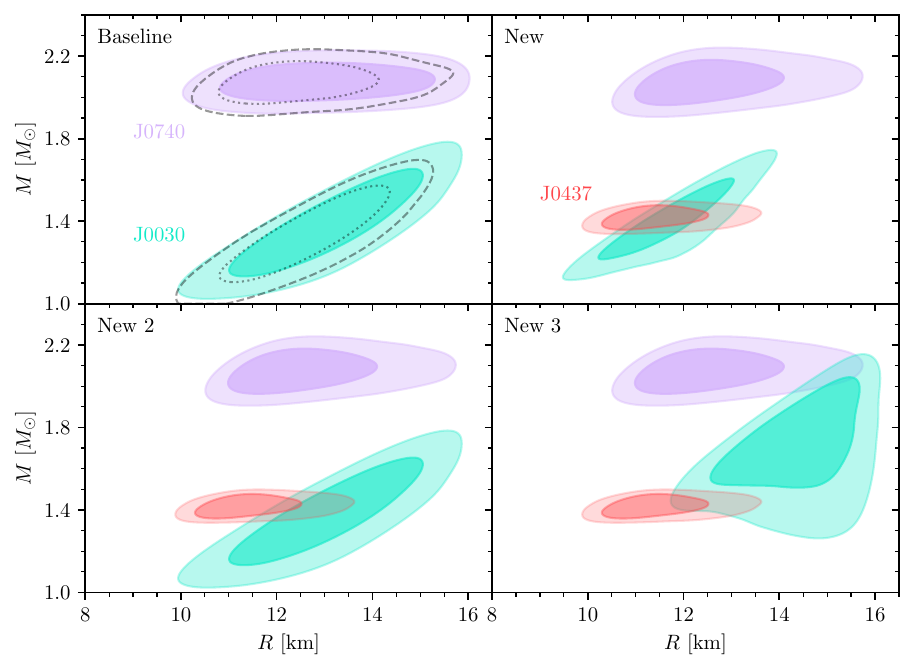}
    \caption{Overview of NICER sources (68\% and 95\% credible regions for mass-radius), for details see Table~\ref{tab:runplan}, with the ``Baseline'' scenario and the three new scenarios with the new J0437 and J0740 NICER results \citep{Choudhury24,Salmi24} and exploring the three possible solutions for J0030 from \citet{Vinciguerra24}. For the ``Baseline'' scenario, we show for comparison the 68\% (95\%) credible regions from \citet{Riley21} and \citet{Riley19} as dotted (dashed) lines, which were used in \citet{Raaijmakers21}.}
    \label{fig:mrdata_appendix}
\end{figure}

\appendix

\section{Results for additional $\chi$EFT bands and astrophysical constraints}

In addition to the results for the \nthreelo\ $\chi$EFT band with the ``Baseline'' and ``New'' scenarios discussed in the main text, we provide in the appendix results for the \citet{Keller2023} \ntwolo\ and the \citet{Hebeler2013} $\chi$EFT band, as well as for two other data scenarios that differ compared to ``New'' only in the J0030 results from \citet{Vinciguerra24}: ``New 2'' is based on \texttt{ST+PST} results for NICER data only, while ``New 3'' is based on \texttt{PDT-U} with a joint analysis of NICER and XMM data. The astrophysical results are summarized in Table~\ref{tab:runplan} and shown as mass-radius regions in Fig.~\ref{fig:mrdata_appendix}. All scenarios include the GW results.

\subsection{Posterior results for the ``New'' scenario and the Hebeler et al.~and \ntwolo\ $\chi$EFT bands}

\begin{figure}[t!]
    \centering
    \includegraphics[width=0.85\linewidth,clip=]{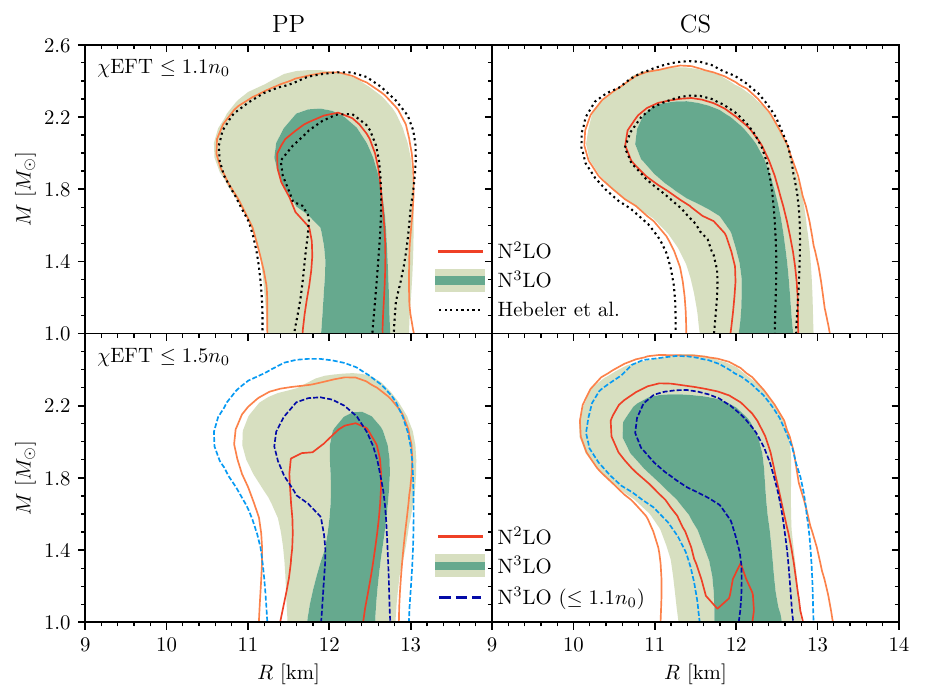}
    \caption{Mass-radius posterior distributions for the PP model (left panels) and CS model (right panels) for the ``New'' scenario. The dark (light) green region and the inner (outer) curves encompass the 68\% (95\%) credible regions of the \nthreelo\ $\chi$EFT band. The top panels compare the posteriors based on the new $\chi$EFT calculations at N$^2$LO (red) and N$^3$LO (green bands) from \citet{Keller2023} to those based on the $\chi$EFT calculations from \citet{Hebeler2013} (dotted black). For the top panels the $\chi$EFT bands are used up to $1.1 n_0$. In the bottom panels, the posterior distributions are shown when using the new $\chi$EFT calculations up to $1.5 n_0$. For comparison, we also show the posterior distribution for N$^3$LO used up to $1.1n_0$ (dashed blue lines).}
    \label{fig:MR_New2_appendix}
\end{figure}

\begin{figure}[t!]
    \centering
    \includegraphics[width=0.85\linewidth,clip=]{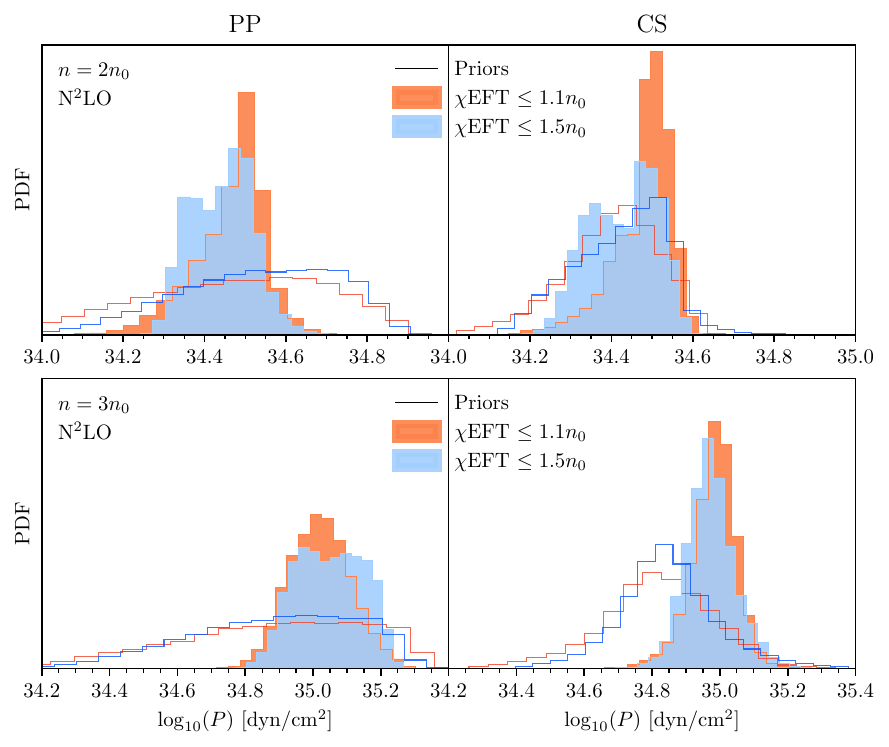}
    \caption{The same as Fig.~\ref{fig:Press_New2} but for N$^2$LO.}
    \label{fig:Press_New2_N2LO}
\end{figure}

In Fig.~\ref{fig:MR_New2_appendix}, we show the mass-radius posteriors for the ``New'' scenario, comparing the previously used \citet{Hebeler2013} band and the \citet{Keller2023} $\chi$EFT bands at \ntwolo\ and \nthreelo\ with transition densities of $1.1 n_0$ and $1.5 n_0$. For the transition density of $1.1 n_0$, the top panels of Fig.~\ref{fig:MR_New2_appendix} show that the mass-radius posteriors of \citet{Hebeler2013} and \ntwolo\ predict similar credible regions to the \nthreelo\ band with only small differences in radii below 1.6\,\Msun. For the transition density of $1.5 n_0$, the \nthreelo\ $\chi$EFT band is also very consistent with the \ntwolo\ band, but at \ntwolo\ we observe a broadening and a hint of a bimodal-like structure more pronounced at \ntwolo\ than for the \nthreelo\ band discussed in the main text. Similar to our results for the \nthreelo\ $\chi$EFT band, the bimodal-like structure also manifests in the posterior distributions of the polytropic fit parameter, $K$, for the \ntwolo\ band as well as for the pressure at intermediate densities of $2 n_0$ and $3 n_0$, as shown in Fig.~\ref{fig:Press_New2_N2LO}.

\begin{figure}[t!]
    \centering
    \includegraphics[width=0.8\linewidth,clip=]{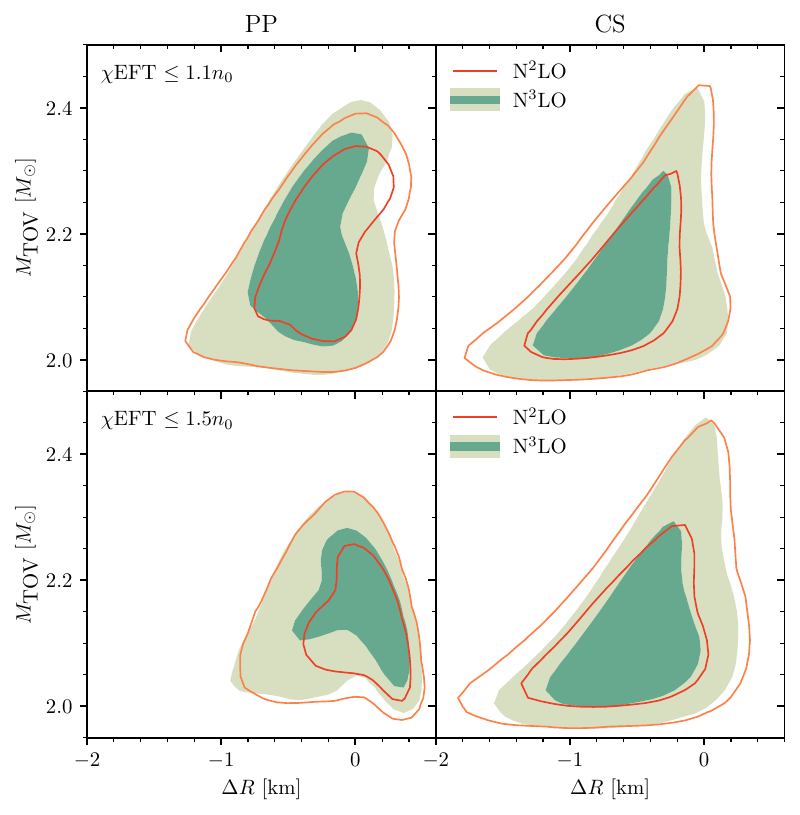}
    \caption{Same as Fig.~\ref{fig:deltaR_maxM_new2} but including also results for the $\chi$EFT bands at \ntwolo.}
\label{fig:deltaR_maxM_new2_appendix}
\end{figure}

As discussed in Sec.~\ref{sec:max_mass}, a key quantity for the dense matter EOS is the maximum mass $M_{\mathrm{TOV}}$. In Fig.~\ref{fig:deltaR_maxM_new2_appendix}, we give the joint posterior distributions of $M_{\mathrm{TOV}}$ and $\Delta R$ for the ``New'' scenario using the \ntwolo\ $\chi$ EFT band up to $1.1 n_0$ and $1.5 n_0$, which are overlaid with the corresponding \nthreelo\ posteriors. This shows that the maximum mass is predicted to be below around 2.4\,\Msun\ (95\% CI) for all $\chi$EFT assumptions, with slightly larger values for the CS models. Moreover, we observe that the posteriors on $M_{\mathrm{TOV}}-\Delta R$ are largely unaffected when going from \ntwolo\ to \nthreelo.

\subsection{Sensitivities to J0030 NICER mass-radius results}

\begin{figure}[t!]
    \centering
    \includegraphics[width=0.85\linewidth,clip=]{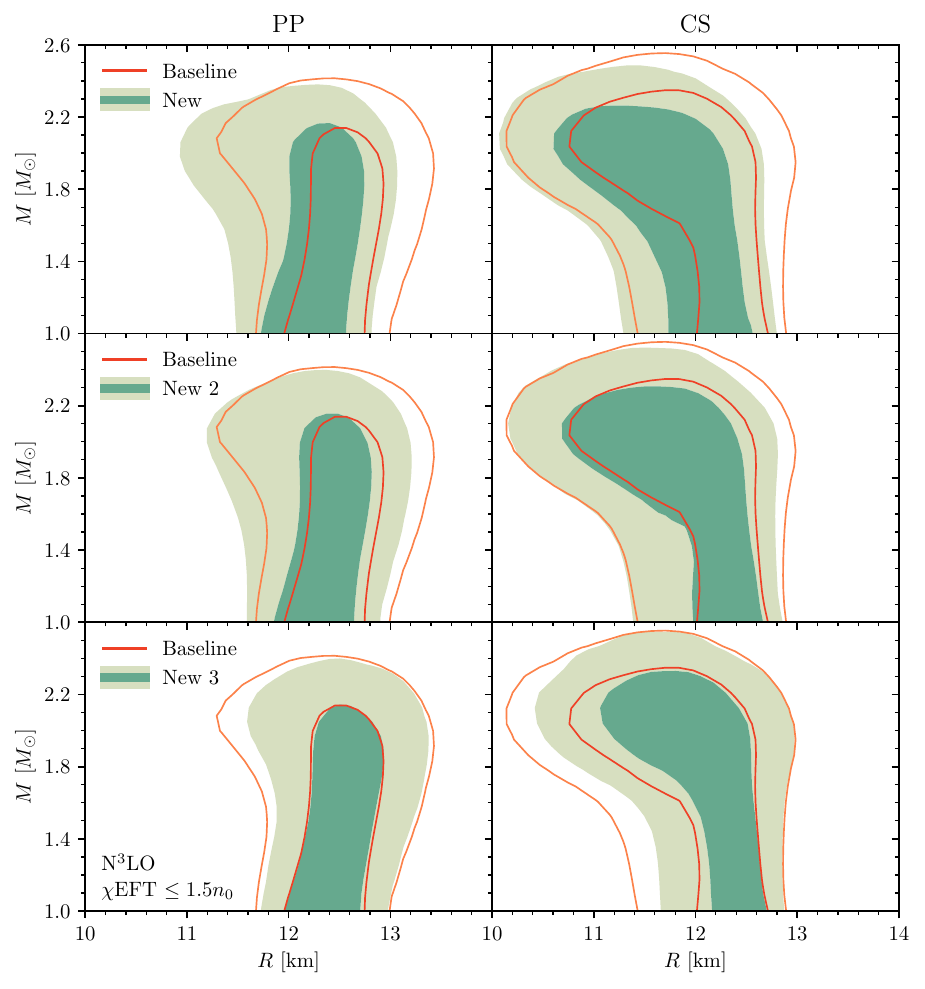}
    \caption{Mass-radius posterior distributions using the \nthreelo\ $\chi$EFT band up to $1.5 n_0$ for the three new NICER scenarios (green shaded regions), in comparison to the ``Baseline'' scenario (red contours), using the PP model (left panels) and the CS model (right panels). The results using the \nthreelo\ $\chi$EFT band up to $1.1 n_0$ are qualitatively very similar.}
    \label{fig:MR_N3LO_15_appendix}
\end{figure}

We next explore the sensitivities to the J0030 NICER mass-radius results by comparing the ``New 2'' and ``New 3'' scenarios to the ``Baseline'' and ``New'' scenarios studied in the main text. To this end, we compare in Fig.~\ref{fig:MR_N3LO_15_appendix} the mass-radius posterior distributions using the \nthreelo\ $\chi$EFT band up to $1.5 n_0$ for the new scenarios to the ``Baseline''. Similar to what we found for ``New,'' the ``New 2'' mass-radius posteriors shift to somewhat lower radii for the PP model, while for the CS model the posteriors narrow. This shifting/narrowing of the ``New 2'' posteriors is due to the combined effect of the J0030 and J0437 NICER results. For the ``New 3'' scenario, the posterior distributions narrow slightly compared to the ``Baseline'' scenario and both the PP and CS models fully remain within the ``Baseline'' contours. In this case, the J0030 \texttt{PDT-U} results, centered around $R=14.44$\,km, compensate for the J0437 result that pushes the posteriors to lower radii, thus yielding mass-radius posteriors that favor neutron stars closer to 12\,km compared to the ``New'' and ``New 2'' scenarios. Depending on the choice of whether to include background constraints for J0030 (and the preferred mode once those constraints are included), the mass-radius posteriors (and, by extension the EOS posteriors) are either---when compared to the ``Baseline''---marginally unaffected (``New 2''), pushed toward lower masses and radii (``New''), or tightened at radii around 12\,km (``New 3''). This variation in the inferred neutron star mass-radius relation highlights the dependence of our inferences on the choice of mass-radius posteriors of J0030.

Finally, in Table.~\ref{tab:ranges_appendix} we list posterior results for the key quantities as in Table.~\ref{tab:ranges} but for all four scenarios, using the \nthreelo\ $\chi$EFT band up to $1.1 n_0$ and $1.5 n_0$ for both the PP and CS models.

\begin{table*}[t!]
\caption{The same as Table.~\ref{tab:ranges}, but with results for the ``New 2'' and ``New 3'' scenarios as well.}
\centering
\begin{tabular}{|c|cccc|cccc|}
\hline
& \multicolumn{4}{c|}{N$^3$LO $\chi$EFT $\le 1.1n_0$} & \multicolumn{4}{c|}{N$^3$LO $\chi$EFT $\le 1.5n_0$} \\ 
& Baseline & New & New 2  & New 3 & Baseline & New & New 2 & New 3 \\
\hline
& \multicolumn{8}{c|}{PP model} \\ 
\hline
$R_{1.4}$ & $12.58^{+0.67}_{-0.75}$ & $12.30^{+0.55}_{-1.04}$ & $12.44^{+0.62}_{-0.95}$ & $12.57^{+0.63}_{-0.67}$  & $12.48^{+0.57}_{-0.67}$ & $12.28^{+0.50}_{-0.76}$ & $12.38^{+0.53}_{-0.73}$ & $12.47^{+0.56}_{-0.55}$ \\
$R_{2.0}$ & $12.40^{+1.12}_{-1.26}$ & $11.99^{+0.93}_{-1.17}$& $12.19^{+1.07}_{-1.19}$ & $12.41^{+1.05}_{-1.04}$ & $12.59^{+0.78}_{-1.24}$ & $12.33^{+0.70}_{-1.34}$ & $12.46^{+0.73}_{-1.29}$ & $12.61^{+0.75}_{-1.02}$\\
$\Delta R$ & $-0.16^{+0.48}_{-0.76}$ & $-0.28^{+0.47}_{-0.71}$& $-0.23^{+0.49}_{-0.69}$& $-0.14^{+0.45}_{-0.68}$ & $0.14^{+0.24}_{-0.77}$ & $0.05^{+0.29}_{-0.78}$ & $0.10^{+0.26}_{-0.75}$ & $0.16^{+0.22}_{-0.67}$\\
\hline
$M_{\rm TOV}$ & $2.27^{+0.16}_{-0.26}$ & $2.15^{+0.20}_{-0.20}$ &$2.20^{+0.18}_{-0.23}$ & $2.26^{+0.16}_{-0.25}$ & $2.17^{+0.15}_{-0.17}$ & $2.15^{+0.14}_{-0.16}$ & $2.16^{+0.14}_{-0.16}$ & $2.16^{+0.15}_{-0.16}$\\
$R_{\rm TOV}$ & $12.03^{+1.51}_{-1.47}$& $11.58^{+1.21}_{-1.21}$& $11.79^{+1.39}_{-1.28}$  & $12.09^{+1.38}_{-1.33}$ & $12.55^{+0.89}_{-1.68}$& $12.22^{+0.83}_{-1.62}$ & $12.40^{+0.83}_{-1.63}$ & $12.58^{+0.84}_{-1.44}$\\
\hline
$\log_{10}$($\varepsilon_{c,\rm{TOV}})$ & $15.13^{+0.26}_{-0.23}$& $15.17^{+0.24}_{-0.20}$& $15.15^{+0.24}_{-0.22}$ & $15.11^{+0.26}_{-0.21}$ & $14.99^{+0.34}_{-0.14}$& $15.04^{+0.31}_{-0.15}$ & $15.01^{+0.31}_{-0.14}$ & $14.98^{+0.31}_{-0.13}$\\
$n_{c,\rm{TOV}}/n_0$ & $4.18^{+2.51}_{-1.50}$ & $4.52^{+2.40}_{-1.45}$ & $4.36^{+2.39}_{-1.49}$ & $4.06^{+2.41}_{-1.37}$ & $3.24^{+2.85}_{-0.79}$ & $3.58^{+2.73}_{-0.90}$ & $3.38^{+2.66}_{-0.82}$ & $3.18^{+2.41}_{-0.73}$\\
$\log_{10}$($P_{c,\rm{TOV}})$ & $35.66^{+0.36}_{-0.35}$& $35.70^{+0.33}_{-0.33}$& $35.68^{+0.33}_{-0.34}$ & $35.64^{+0.35}_{-0.34}$ & $35.44^{+0.48}_{-0.26}$& $35.52^{+0.42}_{-0.26}$ & $35.47^{+0.43}_{-0.25}$ & $35.42^{+0.43}_{-0.23}$\\
\hline
$\log_{10}$($\varepsilon_{c,1.4})$ & $14.87^{+0.11}_{-0.11}$& $14.91^{+0.11}_{-0.09}$& $14.89^{+0.11}_{-0.10}$ & $14.87^{+0.09}_{-0.10}$ & $14.85^{+0.11}_{-0.08}$ & $14.87^{+0.13}_{-0.07}$ & $14.86^{+0.12}_{-0.07}$  & $14.85^{+0.10}_{-0.07}$ \\
$n_{c,1.4}/n_0$ & $2.57^{+0.66}_{-0.54}$ & $2.80^{+0.78}_{-0.48}$ & $2.69^{+0.75}_{-0.53}$ & $2.57^{+0.55}_{-0.51}$ & $2.47^{+0.68}_{-0.38}$ & $2.62^{+0.81}_{-0.37}$ & $2.54^{+0.76}_{-0.37}$ & $2.46^{+0.55}_{-0.36}$\\
$\log_{10}$($P_{c,1.4})$ & $34.96^{+0.15}_{-0.14}$& $35.02^{+0.18}_{-0.11}$& $34.99^{+0.17}_{-0.13}$ & $34.96^{+0.13}_{-0.13}$ & $34.96^{+0.14}_{-0.10}$& $34.99^{+0.17}_{-0.09}$ & $34.97^{+0.16}_{-0.09}$  & $34.96^{+0.12}_{-0.10}$ \\
\hline
$\log_{10}$($\varepsilon_{c,2.0})$ & $15.02^{+0.21}_{-0.17}$& $15.07^{+0.19}_{-0.15}$& $15.05^{+0.19}_{-0.16}$ & $15.01^{+0.19}_{-0.16}$ & $14.95^{+0.22}_{-0.11}$&$14.99^{+0.23}_{-0.11}$ & $14.97^{+0.22}_{-0.10}$ &  $14.95^{+0.19}_{-0.10}$\\
$n_{c,2.0}/n_0$ & $3.43^{+1.75}_{-1.01}$ & $3.85^{+1.69}_{-1.0}$ & $3.65^{+1.62}_{-1.04}$ & $3.40^{+1.48}_{-0.94}$  & $3.01^{+1.66}_{-0.59}$ & $3.27^{+1.86}_{-0.64}$ & $3.13^{+1.73}_{-0.60}$ & $2.99^{+1.37}_{-0.56}$\\
$\log_{10}$($P_{c,2.0})$ & $35.39^{+0.32}_{-0.25}$ & $35.49^{+0.30}_{-0.22}$& $35.44^{+0.30}_{-0.24}$  & $35.39^{+0.27}_{-0.23}$ & $35.31^{+0.32}_{-0.15}$ & $35.37^{+0.35}_{-0.15}$ & $35.34^{+0.33}_{-0.15}$ & $35.31^{+0.27}_{-0.15}$ \\
\hline
& \multicolumn{8}{c|}{CS model} \\ 
\hline
$R_{1.4}$ & $12.44^{+0.41}_{-0.9}$ & $12.29^{+0.47}_{-1.03}$& $12.37^{+0.44}_{-0.86}$  & $12.48^{+0.37}_{-0.62}$  & $12.29^{+0.42}_{-0.94}$ & $12.01^{+0.56}_{-0.75}$ &  $35.34^{+0.33}_{-0.15}$ & $12.34^{+0.38}_{-0.73}$\\
$R_{2.0}$ & $11.91^{+0.8}_{-1.25}$& $11.69^{+0.84}_{-1.12}$& $11.81^{+0.81}_{-1.14}$  & $11.96^{+0.71}_{-0.97}$ & $11.87^{+0.89}_{-1.35}$& $11.55^{+0.94}_{-1.09}$ & $11.76^{+0.86}_{-1.18}$  & $11.98^{+0.80}_{-1.12}$ \\
$\Delta R$ & $-0.52^{+0.52}_{-0.76}$ & $-0.58^{+0.61}_{-0.73}$& $-0.56^{+0.53}_{-0.72}$  & $-0.51^{+0.47}_{-0.70}$ & $-0.40^{+0.60}_{-0.82}$ & $-0.46^{+0.59}_{-0.76}$ & $-0.43^{+0.58}_{-0.77}$ & $-0.36^{+0.55}_{-0.78}$ \\
\hline
$M_{\rm TOV}$ & $2.11^{+0.28}_{-0.16}$ & $2.08^{+0.25}_{-0.17}$ & $2.10^{+0.27}_{-0.16}$ & $2.11^{+0.27}_{-0.16}$ & $2.11^{+0.31}_{-0.16}$ & $2.08^{+0.28}_{-0.16}$ & $2.09^{+0.31}_{-0.17}$ & $2.11^{+0.31}_{-0.17}$ \\ 
$R_{\rm TOV}$ &$11.16^{+1.18}_{-1.15}$& $10.97^{+1.17}_{-1.02}$& $11.06^{+1.16}_{-1.01}$ & $11.29^{+1.04}_{-0.95}$ & $11.25^{+1.38}_{-1.32}$&$10.94^{+1.37}_{-1.04}$ & $11.12^{+1.30}_{-1.16}$ & $11.40^{+1.28}_{-1.12}$ \\
\hline
$\log_{10}$($\varepsilon_{c,\rm{TOV}})$ & $15.38^{+0.09}_{-0.09}$& $15.38^{+0.09}_{-0.09}$& $15.38^{+0.09}_{-0.09}$ & $15.38^{+0.05}_{-0.09}$ & $15.38^{+0.09}_{-0.14}$& $15.38^{+0.09}_{-0.09}$ & $15.38^{+0.09}_{-0.09}$ & $15.38^{+0.09}_{-0.14}$ \\
$n_{c,\rm{TOV}}/n_0$ & $6.53^{+1.41}_{-1.11}$ & $6.64^{+1.35}_{-1.04}$ & $6.58^{+1.37}_{-1.02}$ & $6.49^{+1.16}_{-1.0}$ & $6.65^{+1.51}_{-1.45}$ & $6.82^{+1.44}_{-1.30}$ & $6.70^{+1.51}_{-1.41}$ & $6.56^{+1.58}_{-1.35}$\\
$\log_{10}$($P_{c,\rm{TOV}})$ & $35.89^{+0.24}_{-0.40}$ & $35.90^{+0.24}_{-0.37}$& $35.89^{+0.24}_{-0.38}$ & $35.85^{+0.26}_{-0.36}$ &$35.85^{+0.30}_{-0.46}$& $35.88^{+0.27}_{-0.42}$ & $35.85^{+0.29}_{-0.42}$ & $35.79^{+0.32}_{-0.42}$ \\
\hline
$\log_{10}$($\varepsilon_{c,1.4})$ & $14.90^{+0.12}_{-0.07}$& $14.92^{+0.11}_{-0.07}$& $14.91^{+0.11}_{-0.07}$ & $14.90^{+0.08}_{-0.06}$ & $14.91^{+0.12}_{-0.08}$& $14.94^{+0.10}_{-0.09}$ & $14.92^{+0.11}_{-0.08}$  & $14.90^{+0.10}_{-0.08}$ \\
$n_{c,1.4}/n_0$ & $2.76^{+0.78}_{-0.38}$ & $2.88^{+0.79}_{-0.41}$ & $2.81^{+0.74}_{-0.39}$ & $2.72^{+0.53}_{-0.33}$ & $2.83^{+0.83}_{-0.46}$ & $3.02^{+0.69}_{-0.51}$ & $2.90^{+0.76}_{-0.45}$ & $2.77^{+0.63}_{-0.41}$\\
$\log_{10}$($P_{c,1.4})$ & $35.0^{+0.18}_{-0.09}$& $35.03^{+0.18}_{-0.10}$& $35.01^{+0.17}_{-0.09}$ & $34.99^{+0.12}_{-0.08}$ & $35.02^{+0.18}_{-0.10}$& $35.07^{+0.15}_{-0.11}$ & $35.04^{+0.16}_{-0.09}$ & $35.01^{+0.14}_{-0.09}$ \\
\hline
$\log_{10}$($\varepsilon_{c,2.0})$ & $15.13^{+0.19}_{-0.14}$& $15.16^{+0.18}_{-0.15}$& $15.14^{+0.18}_{-0.14}$ & $15.13^{+0.17}_{-0.13}$ &$15.12^{+0.21}_{-0.17}$& $15.16^{+0.19}_{-0.17}$ & $15.13^{+0.20}_{-0.16}$  & $15.10^{+0.20}_{-0.15}$  \\
$n_{c,2.0}/n_0$ & $4.26^{+1.87}_{-1.06}$ & $4.49^{+1.79}_{-1.11}$ & $4.39^{+1.78}_{-1.08}$ & $4.22^{+1.62}_{-0.97}$ & $4.19^{+2.07}_{-1.18}$ & $4.51^{+1.90}_{-1.26}$ & $4.32^{+1.94}_{-1.18}$ & $4.06^{+1.82}_{-1.05}$\\
$\log_{10}$($P_{c,2.0})$ &$35.54^{+0.33}_{-0.21}$& $35.58^{+0.30}_{-0.21}$& $35.56^{+0.30}_{-0.21}$  & $35.52^{+0.27}_{-0.18}$ & $35.53^{+0.36}_{-0.24}$& $35.59^{+0.31}_{-0.25}$ & $35.55^{+0.32}_{-0.23}$  & $35.50^{+0.31}_{-0.21}$\\
\hline
\end{tabular}
\label{tab:ranges_appendix}
\end{table*}

\clearpage

\bibliography{eos}
\bibliographystyle{aasjournal}

\end{document}